\newif\ifShowKeys
\ShowKeystrue
\ShowKeysfalse

\documentclass[11pt,a4paper]{article} 
\pdfoutput=1
\usepackage[no-natbib-sort]{my-jheppub}

\usepackage{amsmath, amssymb}

\usepackage{bm}
\usepackage{environ}
\usepackage{mathrsfs}
\usepackage{array,arydshln}
\usepackage{booktabs}

\usepackage{graphicx,epsfig}
\usepackage{epic}

\usepackage{tensor} 							

\usepackage{float}
\usepackage[dvipsnames]{xcolor}
\definecolor{maroon}{rgb}{0.8,0.3,0.}

\usepackage{slashed}
\usepackage[nodayofweek]{date time}

\ifShowKeys \usepackage[notcite]{showkeys} \fi

\usepackage{hyperref}

\usepackage{aurical}
\usepackage[T1]{fontenc}

\usepackage{comment}

\usepackage{framed}
\definecolor{shadecolor}{RGB}{255, 230, 204}

\allowdisplaybreaks

\newmuskip\pFqmuskip

\newcommand*\pFq[6][8]{%
  \begingroup 
  \pFqmuskip=#1mu\relax
  \mathcode`\,=\string"8000
  \begingroup\lccode`\~=`\,
  \lowercase{\endgroup\let~}\pFqcomma
  {}_{#2}F_{#3}{\Big[\genfrac..{0pt}{}{#4}{#5};#6\Big]}%
  \endgroup
}

\newcommand*\pFtildeq[6][8]{%
  \begingroup 
  \pFqmuskip=#1mu\relax
  \mathcode`\,=\string"8000
  \begingroup\lccode`\~=`\,
  \lowercase{\endgroup\let~}\pFqcomma
  {}_{#2}\widetilde{F}_{#3}{\Big[\genfrac..{0pt}{}{#4}{#5};#6\Big]}%
  \endgroup
}

\newcommand{\pFqcomma}{\mskip\pFqmuskip}

\newcommand{\be}{\begin{equation}}
\newcommand{\ee}{\end{equation}}

\newcommand{\mc}{\mathcal }

\newcommand{\bv}{\begin{verbatim}}
\newcommand{\ev}{\end{verbatim}}

\newcommand{\la}{\label}

\newcommand{\vp}{\varphi}

\newcommand{\red}[1]{\textcolor{red}{#1}}

\newcommand{\eg}{{\em e.g.}}
\newcommand{\ie}{{\em i.e.}}

\makeatletter
\DeclareFontFamily{OMX}{MnSymbolE}{}
\DeclareSymbolFont{MnLargeSymbols}{OMX}{MnSymbolE}{m}{n}
\SetSymbolFont{MnLargeSymbols}{bold}{OMX}{MnSymbolE}{b}{n}
\DeclareFontShape{OMX}{MnSymbolE}{m}{n}{
    <-6>  MnSymbolE5
   <6-7>  MnSymbolE6
   <7-8>  MnSymbolE7
   <8-9>  MnSymbolE8
   <9-10> MnSymbolE9
  <10-12> MnSymbolE10
  <12->   MnSymbolE12
}{}
\DeclareFontShape{OMX}{MnSymbolE}{b}{n}{
    <-6>  MnSymbolE-Bold5
   <6-7>  MnSymbolE-Bold6
   <7-8>  MnSymbolE-Bold7
   <8-9>  MnSymbolE-Bold8
   <9-10> MnSymbolE-Bold9
  <10-12> MnSymbolE-Bold10
  <12->   MnSymbolE-Bold12
}{}

\let\llangle\@undefined
\let\rrangle\@undefined
\DeclareMathDelimiter{\llangle}{\mathopen}%
                     {MnLargeSymbols}{'164}{MnLargeSymbols}{'164}
\DeclareMathDelimiter{\rrangle}{\mathclose}%
                     {MnLargeSymbols}{'171}{MnLargeSymbols}{'171}
\makeatother

\def \ov {\over}
\def \ci {\cite}
\def \foot {\footnote}
\def \N {{\cal N}}
\def \b{\beta}
\def \m {\mu}
\def \n {\nu}
\def \del{\partial}

\newcommand{\rf}[1]{(\ref{#1})}

\def \k {\kappa}
\def \l {\lambda}
\def \iffa {\iffalse}
 
\def \a  {\alpha}
\def \lan  {\langle}   \def \ran {\rangle}
\def \W {W^{(0)}}
  \def \WW  {W^{(1)}}
 
\def \ve  {\varepsilon}
\def \OO {{\cal O}}
\def \tr {{\rm Tr\,}}
\def \adst {${\rm AdS}_2$ }
\def \g  {\gamma}\def \te {\textstyle} 

\def \sql {{\sqrt \l}}  \def \D   {\Delta}
\def \adss  {AdS$_5 \times S^5\ $}
\def \na {\nabla}\def \G  {\Gamma} \def \ha {\tfrac{1}{2}}
\def \z   {\zeta}

\def \WZ {W^{(\z)}}
\def \lla {\llangle}
\def \rra {\rrangle}
\def \om {\omega}\def \no {\nonumber}
 \def \s  {\sigma}
 \def \P {\Phi}  
\def \t {\tau}

\def \wW  {{\widetilde W}}
\def \wwW {\wW^{(0)}}
  \def \wWW  {\wW^{(1)}}
  \def \wWZ  {\wW^{(\z)} }
   \def \N  {{\cal N}}  \def \zr  {\zeta_{\rm R}}  \def \td {\tilde} 
 \def \C  {{\cal C}}
\def \bb {{\rm b}}
\def \eps {\epsilon}
 \def \G  {\Gamma}
 \def \aa  {{\rm a}}
\def \PP{{\cal P}}
\def \ff {\varkappa} \def \L {\Lambda} \def \kk {\ff} 
\def \TT {{\cal T}}
\def \wa {{{2,1a}}} \def \wb {{{2,1b}}}
\def  \Z  {{\cal Z}}
\def \fF  {{\rm f}}

\title{Non-supersymmetric    Wilson loop   in $\N=4$ SYM  \\
  and   defect   1d CFT   }

\author[a]{Matteo Beccaria,} 

\author[b]{Simone Giombi} 

\author[c]{ and \ \ Arkady A. Tseytlin\footnote{Also at Lebedev Institute, Moscow.}} 

\abstract{ 
Following  
Polchinski and Sully (arXiv:1104.5077),  
we consider  a generalized  Wilson loop  operator containing  
a constant parameter $\z$  in front of the scalar  coupling term, so that 
$\z=0$ corresponds to the standard Wilson loop, while $\z=1$ to the 
locally supersymmetric one.  We compute the expectation value 
of this operator for circular loop as a function of $\z$ to second order in the planar weak coupling expansion in $\N=4$ SYM theory. 
We then explain the relation of the expansion near the two conformal points $\z=0$ and $\z=1$  
to the correlators of  scalar operators inserted on the loop. 
We also discuss  the \adss  string 1-loop  correction to the strong-coupling expansion of the standard circular Wilson loop, 
as well as its generalization to the case of mixed boundary conditions on the five-sphere coordinates, corresponding to general $\z$.
From the point of view of the defect CFT$_1$   defined on the Wilson line, the  
$\z$-dependent term can be seen as a perturbation 
driving a RG flow from the standard Wilson loop in the UV to the supersymmetric Wilson loop in the IR. 
Both at weak and strong coupling we find that  the logarithm of the expectation value of the standard Wilson loop
for the circular contour is larger than that of the supersymmetric one, which appears to be in agreement with the 1d analog of the F-theorem. 
}

\affiliation[a]{Dipartimento di Matematica e Fisica Ennio De Giorgi,\\
Universit\`a del Salento \& INFN, Via Arnesano, 73100 Lecce, 
Italy} 

\affiliation[b]{Department of Physics, Princeton University, Princeton, NJ 08544, USA}

\affiliation[c]{Blackett Laboratory, Imperial College, London SW7 2AZ, U.K.}
                       
\emailAdd{matteo.beccaria@le.infn.it, sgiombi@princeton.edu, tseytlin@imperial.ac.uk} 

\begin{document}
\date{\currenttime}

\begin{flushright}\small{ PUPT-2543\ \  \ \ \ \ \ \ \  \ \   \ \ \ \ \ \ \ \ \\   NSF-ITP-17-141\ \ \ \ \  \ \ \ \ \ \ \ \ \\ Imperial-TP-AT-2017-{09}\\ }\end{flushright}				
\maketitle
\flushbottom

\iffa 

Title:  Non-supersymmetric    Wilson loop   in N=4  SYM    and   defect   1d CFT
Authors:    Matteo  Beccaria, Simone Giombi,   Arkady  Tseytlin
Abstract:
Following  arXiv:1104.5077,  we consider  a generalized  Wilson loop  operator containing  
a constant parameter $\zeta$  in front of the scalar  coupling term, so that  $\zeta=0$ corresponds to the standard Wilson loop, while $\zeta=1$ to the  locally supersymmetric one.  We compute the expectation value  of this operator for circular loop as a function of $\zeta$ to second order in the planar weak coupling expansion in N=4 SYM theory.  We then explain the relation of the expansion near the two conformal points $\zeta=0$ and $\zeta=1$  to the correlators of  scalar operators inserted on the loop.  We also discuss  the \adss  string 1-loop  correction to the strong-coupling expansion of the standard circular Wilson loop,  as well as its generalization to the case of mixed boundary conditions on the five-sphere coordinates, corresponding to general $\zeta$. From the point of view of the  defect    1d  CFT defined on the Wilson line, the    $\zeta$-dependent term can be seen as a perturbation driving a RG flow from the standard Wilson loop in the UV to the supersymmetric Wilson loop in the IR.  Both at weak and strong coupling we find that  the logarithm of the expectation value of the standard Wilson loop for the circular contour is larger than that of the supersymmetric one, which appears to be in agreement with the 1d analog of the
 F-theorem. 

Comments:   36 pages
Report Number:  PUPT-2543, NSF-ITP-17-141,  Imperial-TP-AAT-2017-09

\fi

\section{Introduction}
The   expectation value of the 
Wilson loop  (WL)  operator $\lan  \tr\PP e^{i\int A}\ran $    is an important observable in any gauge theory.
In  $\N=4$ Super Yang-Mills (SYM),  the  Wilson-Maldacena loop (WML)  \cite{Maldacena:1998im, Rey:1998ik}, 
which contains  an extra scalar coupling making it  locally-supersymmetric,    
was  at the center of attention, but  
 the  study of  the  ordinary, ``non-supersymmetric''  WL is  also  of interest   \cite{Alday:2007he,Polchinski:2011im} in the context of 
the AdS/CFT correspondence. 
Computing the large $N$  expectation value  of the standard   WL  
for some simple contours  (like circle or cusp)  should produce new non-trivial  functions  of the  't Hooft  coupling $\l= g^2 N$  
which are no longer controlled  by supersymmetry  but may still be  possible to determine  using  
the underlying  integrability of the theory. 
Another motivation  comes from considering correlation functions of  local operators  inserted along the WL:
this should produce a new example of AdS$_{2}$/CFT$_{1}$  duality, 
similar but different from the one recently discussed in the WML  case  \ci{Cooke:2017qgm,Giombi:2017cqn}. 
In the latter case, correlators of local operators on the 1/2-BPS Wilson line have a $OSp(4^*|4)$ 1d superconformal symmetry, 
while in the ordinary WL case one expects a non-supersymmetric ``defect'' CFT$_1$ with $SO(3)\times SO(6)$ ``internal'' 
symmetry.

On general grounds, for the  standard WL defined for a smooth contour  one should find that 
  (i)  all  power  divergences (that  cancel in the WML  case)   
exponentiate and factorize  \cite{Polyakov:1980ca,Dotsenko:1979wb,Gervais:1979fv,Arefeva:1980zd,Dorn:1986dt,Marinho:1987fs}  and (ii) 
 all logarithmic divergences     cancel  as the gauge  coupling  is not running in $\N=4$  SYM theory. Thus 
   its   large $N$ expectation value   should  produce   a nontrivial  finite function of $\l$
   (after  factorising power  divergences, or directly, if computed  in dimensional regularization). 
 
 It is   useful to  consider a 1-parameter  family of  Wilson loop operators  with an arbitrary coefficient $\z$ in front of the scalar 
 coupling   which  interpolates     between  the WL  ($\z=0$) and the WML ($\z=1$)  cases 
  \ci{Polchinski:2011im}  
\be
\la{0}
W^{(\z)}  (C) = \frac{1}{N}\,\text{Tr}\,\mc P\,\exp\oint_{C}d\tau\,\big[i\,A_{\mu}(x)\,\dot x^{\mu}
+ \z \Phi_{m}(x) \, \theta^{m} \,|\dot x|\,\big], \qquad\qquad  \theta_{m}^{2}=1\ . 
\ee
We   may     choose   the  direction $\theta_m$  of  the 
scalar  coupling  in \rf{0}  to be along 6-th direction, i.e. $\P_m \theta^m =\P_6$.
 Below   we shall   sometimes omit   the  expectation value    brackets using  the notation  
\be  \la{1111} 
{\rm WL}: \ \ \ \ \  \lan W^{(0)} \ran \equiv \W \ , \ \ \ \qquad\qquad   \ \ \    {\rm WML}: \ \ \ \ \   \lan W^{(1)}\ran 
\equiv \WW \ . \ee
Ignoring power divergences, for generic $\z$  the expectation value
 $\lan \WZ \ran$  for a smooth contour   may have  additional logarithmic 
divergences  but it should be  possible to  absorb  them  into a renormalization of  the coupling $\z$,  \ie \foot{Here there is an
 analogy  with a 
 partition function of a renormalizable QFT:    if $g_\bb$   is  bare coupling  depending on cutoff $\L$ 
one has $Z_\bb ( g_\bb(\L), \L) = Z (g(\m), \m)$, \ \ 
$\mu {dZ\ov d  \mu} =\m { \del Z \ov \del  \m}  + \beta (g) {\del Z \ov \del g} =0$, \ \ $\beta =\m {d g \ov d  \m}$.
In the present case the expectation value depends on $\mu$  via $\mu R$   where $R$ is  the radius of the circle (which we often set to 1). 
A natural  choice of  renormalization point is then $\mu=R^{-1}$.
}
\be \la{113}
\lan W^{(\z)}\ran  \equiv W\big(\l; \z(\mu), \mu\big)   
 \ , \ \ \ \  \qquad   \mu {\del \ov \del \mu}  W  + \beta_\z  { \del \ov \del \z} W =0  \ , 
\ee
 where   $\mu$ is  a  renormalization scale and  the beta-function  is, to leading order at weak coupling 
 \ci{Polchinski:2011im} 
 \be  
\beta_\z =\mu  { d \z \ov d  \mu } = - {\l \ov 8 \pi^2}  \z (1-\z^2)   + \mc O  ( \l^2)   \ .  \la{111} \ee
The WL and WML  cases  in \rf{1111} 
are the  two  conformal   fixed points $\z=0$  and $\z=1$ 
where  the   logarithmic divergences     cancel out  automatically.\foot{As 
 the expectation value of the standard WL    has no logarithmic divergences, 
combined   with the fact that  the  straight line (or circle) preserves  a subgroup of 4d conformal group 
this implies that  one  should   have   a 1d conformal  $SL(2,R)$ 
 invariance for the  corresponding  CFT on the line for all  $\lambda$.} 
 Given that the   SYM action is invariant  under the change of sign of $\Phi_m$  the fixed point points
 $\z=\pm 1$ are equivalent  (we may resstrict $\z$  to be non-negative in \rf{0}).

 Our   aim  below  will    be to compute the  leading weak   and strong  coupling terms 
 in the   WL expectation  value   for a  circular contour in the planar limit. 
 As is well known,  the  circular WML  expectation  value  can be   found  exactly 
 due to underlying supersymmetry;  in the planar limit  \cite{Erickson:2000af,Drukker:2000rr,Pestun:2007rz} 
 (see also 
 \ci{Zarembo:2016bbk})
 \be
\la{1.1}
 \WW (\text{circle}) = \frac{2}{\sqrt\lambda}\,I_{1}(\sqrt\lambda) = \begin{cases}
1+\frac{\lambda}{8}+\frac{\lambda^{2}}{192}+\cdots, &\qquad  \lambda\ll 1\ , \\
\sqrt{\frac{2}{\pi}}\, { {1\ov ( \sqrt \lambda)^{3/2} } }\, e^{\sqrt\lambda}\,\big(1-\frac{3}{8\,\sqrt\lambda}+\cdots\big), & \qquad \lambda \gg 1\ .
\end{cases}
\ee
For a straight line the expectation value of the  WML  is 1, and  then  for  the circle  its  non-trivial  value  can be understood 
as  a consequence of an anomaly in the conformal transformation  relating the line to the circle 
\cite{Drukker:2000rr}. As this  anomaly is  due to  an  IR  behaviour of the vector field   propagator 
\cite{Drukker:2000rr}, one may  wonder   if the same anomaly argument may apply to the 
  WL    as well. Indeed, in this case  ($\z=0$)   there are no additional logarithmic divergences 
  and then  after all power divergences are  factorized or   regularized away   one   gets  
  $ \W (\text{line})  =1$; then the finite part   of $ \W (\text{circle}) $ 
      may happen  to be  the same as in the WML  case \rf{1.1}.\foot{The conjecture  that   the circular  WL   may have 
         the  same value as the locally-supersymmetric    WML one 
  runs, of course,  against the  derivation of the expectation value of the latter based on localization 
 \ci{Pestun:2007rz} as there  is no reason why the  localisation  argument should apply in the  standard WL case.}

Some  indication  in favour of this is that 
  the leading strong and weak coupling terms in the   circular  WL   happen to be     the same as in  the WML   case. 
 The leading strong-coupling term  is  determined by 
 the volume of the same minimal surface  (\adst   with  circle as a boundary)   given by  
$2 \pi ( {1\ov a} - 1) $  and (after subtracting the  linear divergence) 
  thus   has  the   universal   form   $ \lan \WZ\ran \sim e^{ \sqrt \lambda}$.
At weak coupling,  the circular WL and WML   also   have the same
 leading-order expectation value  (again after subtracting linear divergence) 
$\lan \WZ \ran= 1 + {1\ov 8}\l +  O(\l^2)$.  

However,  as we shall  see  below,  the subleading terms  in WL  in  both weak and strong  coupling 
expansion   start  to differ from the WML  values, \ie\,   $\lan \WZ(\rm circle)\ran$  develops dependence on $\z$. 
This   implies, in particular,    that  the  conformal anomaly argument of \cite{Drukker:2000rr}   does not apply for $\z =0$.\foot{This may be attributed to the 
   presence of  extra (power)  divergences  that do not cancel  automatically   in the standard WL case.   For generic $\z$  there 
are also additional logarithmic divergences that break conformal invariance.}

 Explicitly, we shall   find  that at  weak coupling  (in dimensional regularization)
 \begin{align} &\la{1}
\lan  \WZ \ran  = 1 +  {1\ov 8} \l  +  \Big[ \frac{1}{192}+\frac{1}{128\,\pi^{2}} (1-\z^2)^2 \Big]\l^2  +  \OO(\l^3) \ . 
\end{align}
This interpolates between the WML value in \rf{1.1} and  the WL value ($\zeta=0$)
 \be \la{91}  \W= 1 +  {1\ov 8} \l  +  
 \Big( \frac{1}{192}+\frac{1}{128\,\pi^{2}} \Big)\l^2  +  \OO(\l^3)  \ . \ee
 Note that the  2-loop  correction  in \rf{91} to the WML  value in  \rf{1.1} 
 has a different transcendentality; it would  be very interesting to  find  the all-order    generalization of \rf{91}, i.e. 
the  counterpart of the  exact  Bessel function expression 
 in \rf{1.1} in  the   standard WL  case. It is tempting to conjecture that the highest transcendentality part of 
$\lan W \ran$ at each order in the perturbative expansion is the same for supersymmetric and non-supersymmetric Wilson loops and  hence given by \rf{1.1}.  

 The  expression   \rf{1} passes    several  consistency checks.  The UV finiteness  of the two-loop $\l^2$ term is 
 in agreement  with $\z$-independence of the one-loop term  (cf. \rf{113},\rf{111}  implying that 
 UV logs  should appear   first at the next $\l^3$ order).  
 The derivative of  (log of)  \rf{1} 
 over $\z$ is proportional to the beta-function \rf{111}
 \be \la{2}
 {\del \ov \del \z } \log  \lan \WZ \ran =   \C \, \beta_\z  \ , \ \ \qquad \qquad    \C= {\l\ov 4}  +   \OO(\l^2) \ , 
 \ee
 where $\C=\C(\l,\z)$   should not have zeroes. 
 This implies that  the conformal points $\z=1$  and $\z=0$   are  extrema  (minimum  and maximum)   of 
 $\lan \WZ \ran $. 
 This is   consistent with  the interpretation of $\lan \WZ \ran $ as a 1d  partition function  on $S^1$ 
  that may be computed in  conformal perturbation theory   near $\z=1$  or $\z=0$    conformal points.
  Indeed, eq.\rf{2}   may be viewed as a special $d=1$ case of the  relation   
   ${\del F\ov  \del g_i}  = \C^{ij} \beta_j $    for free   energy $F$ on a sphere $S^d$ 
  computed by perturbing a CFT$_d$  by  a  linear combination of  operators $g_i  O^i$
   (see, e.g.,  \ci{Klebanov:2011gs,Fei:2015oha}). 
   
In the present case, the flow   \ci{Polchinski:2011im}  
 is driven by the scalar operator $\P_m \theta^m =\P_6$  in \rf{0}   restricted to the line, 
and the condition  ${\del \ov \del \z } \lan \WZ \ran \big|_{\z=0, 1}  =0$  means that its 
one-point function vanishes at the conformal points, as required  by the 1d conformal invariance. 
The parameter $\zeta$ 
may be viewed as a ``weakly relevant''  (nearly marginal  up to  $\OO(\l)$  term, cf. \rf{111})  
coupling constant    
 running from $\zeta=0$ in the UV (the ordinary Wilson loop) 
to $\zeta=1$ in the IR (the supersymmetric Wilson loop). Note that our  result (\ref{1}) implies that 
\be \la{190}
\log \lan W^{(0)} \ran\  >\  \log \lan W^{(1)} \ran \ . 
\ee
Hence, viewing $\langle \WZ \rangle = Z_{S^1} $ as a partition function of a 1d  QFT on the circle, 
this is precisely consistent
 with the $F$-theorem \cite{Myers:2010xs, Klebanov:2011gs, Casini:2012ei, Giombi:2014xxa,Fei:2015oha}, which in $d=1$
(where it is  analogous to 
  the $g$-theorem \ci{Affleck:1991tk,Friedan:2003yc} applying to the boundary of a 2d theory) 
implies 
\be \la{110} 
\widetilde F_{_{\rm UV}} > \widetilde F_{_{\rm IR}}\,,
\qquad\qquad  \widetilde F \equiv  {\te \sin { \pi d \ov 2} }\, \log Z_{S^d} \Big|_{d=1}  = \log Z_{S^1} =-F\,.
\ee 
Moreover, we see that $ \lan W^{(\zeta)}\ran$ decreases monotonically as a function of $\zeta$ 
from the non-supersymmetric to the supersymmetric fixed point. 

The second derivative of $\lan \WZ \ran$    which from \rf{2} 
is thus  proportional to   the derivative of the beta-function \rf{111}
 \be \la{4}
 {\del^2 \ov \del \z^2 }\log  \lan \WZ \ran \Big|_{\z=0, 1}  =   \C\,  {\del \beta_{\z}  \ov \del \z} \Big|_{\z=0, 1} 
 \ , \ee
 should, on the other hand,   be given by the integrated  2-point function  of $\P_6$ restricted to the line  and   should 
 thus be  determined   by the  corresponding anomalous   dimensions.
 Indeed,   $ {\del \beta_{\z}  \ov \del \z} \big|_{\z=0, 1} $   
  reproduces  \ci{Polchinski:2011im} the anomalous dimensions \ci{Alday:2007he}    of $\P_6$ at   the $\z=1$ and $\z=0$   conformal points
  \be
\begin{aligned} 
\la{3}  
 &\Delta(\z) -1=  {\del \beta_\z  \ov \del \z} = {\l\ov 8\pi^2} ( 3 \z^2 -1)   + \OO(\l^2)\ ,  \\ 
 &\Delta(1) = 1 + {\l \ov  4 \pi^2} + \ldots \ , \qquad \ \ \ \ \  \ \  \Delta(0)  = 1 - {\l \ov  8 \pi^2} + \ldots  \ .
\end{aligned}
 \ee
 Again, this is a   special   case of a general  relation   between the second derivative of free energy  on $S^d$ at conformal 
 point and anomalous dimensions found in conformal perturbation theory. 
 We shall  explicitly   verify  this   relation between ${\del^2 \ov \del \z^2 } \lan \WZ \ran \big|_{\z=0, 1}$   and  the 
  integrated 2-point function of   $\P_6$  inserted into the  circular Wilson  loop         in section \ref{sec4} below.

The interpretation  of $\langle \WZ \rangle$  
as a partition  function  of an effective 1d QFT  is strongly supported by its  strong-coupling    representation as  
the  \adss string theory   partition function on a disc  with  mixed boundary   conditions \ci{Polchinski:2011im}
for  $S^5$  coordinates  (in particular, Dirichlet for $\z=1$ and Neumann  for $\z=0$  \ci{Alday:2007he}). 
As we will   find   in section \ref{secstr},   
 in contrast to the  large $\l$ asymptotics  
 of the  WML  $ \langle \WW \rangle \sim  (\sql)^{-3/2} e^{\sql} + ... $ in \rf{1.1}, in the  standard WL 
 one gets    \be\la{03}  \langle \W \rangle \sim  \sql \,  e^{\sql} + ...  \  , \ee
 so that  the F-theorem inequality \rf{190},\rf{110}  is satisfied also at strong coupling. 
 At strong coupling,  the counterpart of the $\P_6$  perturbation  near the $\z=0$ conformal point 
 is an extra  boundary  term  (which to leading order  is quadratic  in $S^5$  coordinates)
    added to  the string   action  with Neumann  boundary condition  to   induce   the  boundary RG flow to the 
    other conformal point. \footnote{
    In particular, the boundary term is independent of the fermionic fields. When restricted to 
    the AdS$_{2}$ minimal surface dual to the Wilson loop, they will be assumed to have
    the usual unitary $\Delta=3/2$ boundary behaviour along the whole RG flow. Instead, 
    for the $S^{5}$ scalars the 
    boundary deformation induces  unitary 
    mixed boundary conditions and only in Dirichlet case 
    we have unbroken supersymmetry.
    }
        The    counterpart of $\z$   in \rf{0}   is a 
     (relevant) coupling $\kk=  {\fF} (\z; \l)$ 
    (which is 0 for $\z=0$ and $\infty$ for $\z=1$)  has  the beta function (see \ref{sec4.2}) 
    $\beta_\kk= ( - 1 + {5\ov \sql}) \kk + ...$. This  
      implies  that  strong-coupling dimensions   of $\P_6$  near  the two conformal points  should be   (in agreement with 
      \ci{Alday:2007he,Giombi:2017cqn}) 
      \be 
      \Delta-1 = \pm \big( - 1 + {5\ov \sql}  + ... \big) \ , \ \ \   {\rm i.e.} \ \ \ \ 
      \Delta(0) = {5\ov \sql}  + ...   \ , \ \ \ \    \Delta(1) = 2 - {5\ov \sql}  + ...  \ . \la{04} \ee

This  paper is organized as follows.  In section \ref{sec2}     we shall   compute the two leading   terms  in the planar  weak-coupling expansion   of the    circular  WL. 
The structure of the   computation   will be  similar to the one
in  the WML  case   in \cite{Erickson:2000af} (see also \ci{Young:2003pra})
 but now the integrands  (and thus  evaluating   the   resulting  path-ordered integrals)
   will be  substantially  more complicated. 
  We shall   then generalize to  any value of $\z$ in \rf{0}  obtaining the expression in \rf{1}. 
  
In section \ref{sec4} we shall  elaborate on the  relation   between  the  expansion of the generalized WL \rf{1} 
near the  conformal     points   and the  correlators of   scalar operators  inserted on the loop. 
In section \ref{secstr}   we shall consider  the  strong-coupling (string theory) computation
of the  circular  WL   to 1-loop   order in \adss   superstring theory 
generalizing the previous discussions in  the WML  case.  We shall  also discuss the general $\z$  case 
in section \ref{sec4.2}.
   
Some concluding remarks will be made in section \ref{secf}.
In Appendix \ref{A}    we shall  comment on    cutoff  regularization. 
In Appendix \ref{B}      we shall  explain   different methods of computing path-ordered  integrals on a circle appearing 
in the  2-loop ladder diagram  contribution to the generalized WL.

\section{Weak coupling expansion}\label{sec2} 

Let us now  consider  the weak-coupling ($\l=g^2 N \ll 1$) expansion   in planar $\N=4$  SYM    theory 
and   compute the  first two leading terms  in the expectation  value for 
the generalized  circular Wilson  loop \rf{0} 
\be \la{3.3}
\langle \WZ\rangle = 1+\ \l\, \WZ_{1}+\ \l^{2}\, \WZ_{2}+\cdots .
\ee
We shall   first  discuss  explicitly the   standard Wilson loop $\W$ in \rf{1111} 
comparing  it  to the    Wilson-Maldacena   loop  $\WW$  case    in \cite{Erickson:2000af}
and then   generalize to an arbitrary  value of  the parameter $\z$.

\subsection{One-loop  order}

\begin{figure}[t]  
\centering
\includegraphics[scale=0.25]{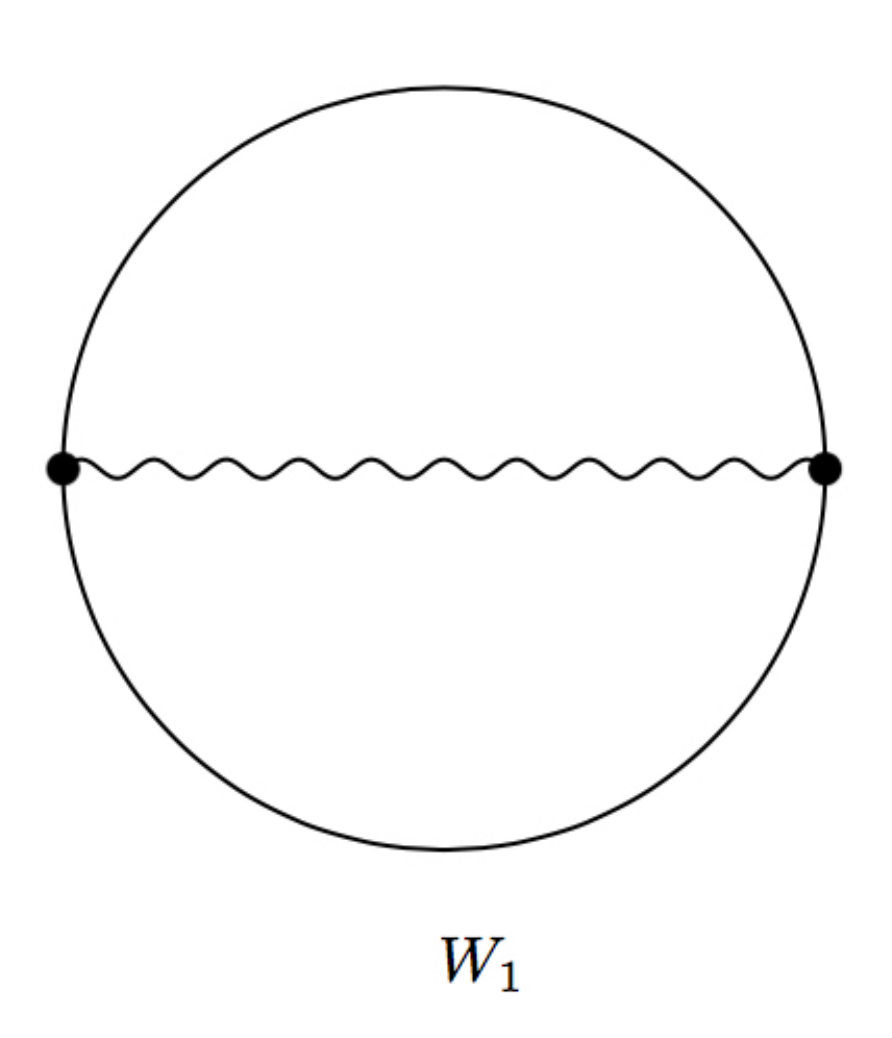}
\caption[] {
Gauge field exchange diagram 
contributing the standard  Wilson loop at the  leading order.
In the Wilson-Maldacena loop case there is an additional scalar  exchange contribution. 
\label{fig:one-loop}}
\end{figure}

The perturbative computation  of the   WML   was  discussed in 
\cite{Erickson:2000af} (see also   \cite{Young:2003pra})   that we shall follow  and generalize.
 The  order $\l$  contribution is\footnote{There is a misprint in the overall coefficient in  \cite{Erickson:2000af} corrected  in \cite{Young:2003pra}. 
}
\be\la{3.4}
\WW_{1}(C) = \frac{1}{(4\,\pi)^{2}}\oint_{C} d\tau_{1}d\tau_{2}\ 
\frac{|\dot x(\tau_{1})|\,|\dot x(\tau_{2})|-\dot x(\tau_{1})\cdot \dot x(\tau_{2})}{
|x(\tau_{1})-x(\tau_{2})|^{2}}  \ . 
\ee
Here  the term $\dot x(\tau_{1})\cdot \dot x(\tau_{2})$   comes from the vector exchange (see  Fig. \ref{fig:one-loop}) 
and the  term  $|\dot x(\tau_{1})|\,|\dot x(\tau_{2})|$   from the scalar exchange. 
This integral  is finite for a smooth loop.
 In particular, for the straight  line
$x^\m(\tau) = (\tau,0,0,0)$,
the numerator in $\WW_{1}$ is zero and thus 
\be
\la{3.5}
\WW_{1}(\text{line})=0.
\ee
For the circular loop, $x^\m(\tau) = (\cos\tau, \sin\tau, 0, 0)$,
the integrand in \rf{3.4}  is  constant
 \be\la{399}
\frac{|\dot x(\tau_1)|\,|\dot x(\tau_2)|-\dot x(\tau_1)\cdot \dot x(\tau_2)}{
|x(\tau_1)-x(\tau_2)|^{2}} = \frac{1}{2}
\ee
and thus, in  agreement with \rf{1.1},\rf{3.3} 
\be
\la{3.6}
\WW_{1}(\text{circle}) = \frac{1}{(4\pi)^{2}}\,(2\pi)^{2}\,\frac{1}{2} = \frac{1}{8} \ .
\ee
 %
The  analog of \rf{3.4}  in the case of the standard WL is   found  by  omitting the 
 scalar exchange  $|\dot x(\tau_{1})|\,|\dot x(\tau_{2})|$   term  in the integrand.
 The resulting integral will  have linear divergence  (see  Appendix \ref{A})
 that can be 
 factorized or automatically ignored    using dimension  regularization
 for the vector propagator  with parameter $\om=2-\ve  \to 2$. If we   replace the dimension $4$  by $d=2 \omega\equiv  4 -2 \ve$
  the   standard  Euclidean  4d  propagator   becomes
\be\la{3.7}
\Delta(x) = (-\del^2)^{-1} = \frac{\Gamma(\omega-1)}{4\pi^{\omega}} { 1 \ov |x|^{2 \om -2}}
\ . \ee
Then 
\be\la{3.8}
\W_{1} = \frac{1}{(4\,\pi)^{2}}\oint d\tau_{1}d\tau_{2}\,
\frac{-\dot x(\tau_{1})\cdot \dot x(\tau_{2})}{
|x(\tau_{1})-x(\tau_{2})|^{2}}\  \to \ 
\frac{\G(\om-1)}{16\pi^{\omega}}\oint d\tau_{1}d\tau_{2}\,
\frac{-\dot x(\tau_{1})\cdot \dot x(\tau_{2})}{
|x(\tau_{1})-x(\tau_{2})|^{2\omega-2}}\,.
\ee
In the infinite  line case       we  get ($L\to \infty$) \footnote{We use that  $\int_{0}^{L}d\tau_{1}\int_{0}^{L}d\tau_{2}\,f(|\tau_{1}-\tau_{2}|)
=2\,\int_{0}^{L}d\tau\,\,(L-\tau)\,f(\tau)$.}
\begin{align}
&\int_{0}^{L} d\tau_{1}\int_{0}^{L}d\tau_{2}\,\frac{1}
 {|\tau_{1}-\tau_{2}|^{2\omega-2}} =
2\, \int_{0}^{L} d\tau\,\frac{L-\tau}{\tau^{2\,(\omega-1)}} = 
 \frac{L^{4-2\omega}}
 {2-\omega}\frac{1}{3-2\omega} \to 0 \ .  \la{3.9}
 \end{align}
%
 The formal  integral here is   linearly divergent. 
 If  we  use dimensional regularization to regulate both 
 UV and IR   divergences     (analytically continuing from $\om > 2  $ region)   we get   
   as in \rf{3.5} 
 \be\la{3.10}
\W_{1}(\text{line})=0.
\ee
 In the case of a circle, we   may use \rf{3.4},\rf{3.6} to  write ($\omega \equiv 2-\varepsilon\to 2$)
 \begin{align}
 \la{3.11}
\W_{1}(\text{circle}) &= 
\WW_{1}(\text{circle})
-\frac{ \G(\om-1)  }{16\pi^{\omega}}\oint d\tau_{1}d\tau_{2}\,
\frac{1}{
|x(\tau_{1})-x(\tau_{2})|^{2\omega-2}}\notag \\
&=
\frac{1}{8}
-\frac{\G(\om-1)  }{2^{2\omega+2}\pi^{\omega}}\oint d\tau_{1}d\tau_{2}\,\Big[
\sin^{2} \tfrac{\tau_{12}}{2}\Big]^{1-\omega} \ . 
 \end{align}
 The integral  here may be computed, e.g., by  using the master-integral in 
 eq.(G.1) of \cite{Bianchi:2016vvm}\foot{Alternative  direct methods  of computing similar integrals 
 are discussed in Appendix \ref{B}. We also note that  such  2-point and 3-point integrals 
 can 
be viewed as a special $d=1$ case of the conformal integrals on $S^d$ used in \cite{Cardy:1988cwa,Klebanov:2011gs}.}
\begin{align}
\la{3.12}
\mc M(a,b,c) &\equiv  \oint d{\tau_1}d{\tau_2}d{\tau_3} \Big[\sin^{2}\tfrac{\tau_{12}}{2}\Big]^{a}
\Big[\sin^{2}\tfrac{\tau_{23}}{2}\Big]^{b}
\Big[\sin^{2}\tfrac{\tau_{13}}{2}\Big]^{c}\notag \\
&= 8\,\pi^{3/2}\,
\frac{
\Gamma(\frac{1}{2}+a)\Gamma(\frac{1}{2}+b)\Gamma(\frac{1}{2}+c)\Gamma(1+a+b+c)}{
\Gamma(1+a+c)\Gamma(1+b+c)\Gamma(1+a+b)} \ , 
\end{align}
i.e. 
\be\la{3331}
\oint d\tau_{1}d\tau_{2}\,\Big[
\sin^{2}\tfrac{\tau_{12}}{2}\Big]^{1-\omega} = \frac{1}{2\pi}\,\mc M(1-\omega,0,0)
= \frac{4\,\pi^{3/2}\,\Gamma(-\frac{1}{2}+\varepsilon)}{\Gamma(\varepsilon)} = - 8 \pi^2 \ve + \mc O(\varepsilon^2).
\ee
Plugging this into (\ref{3.11}), we  get  the same result as in   (\ref{3.6}):
\be\la{3466}
\W_{1}(\text{circle}) = \frac{1}{8} \ .
\ee
Thus   the leading-order   expectation values for 
 the WML  and WL  are the same   for both  the straight line and the circle.  

\subsection{Two-loop order } 

\begin{figure}[t]  
\centering
\includegraphics[scale=0.3]{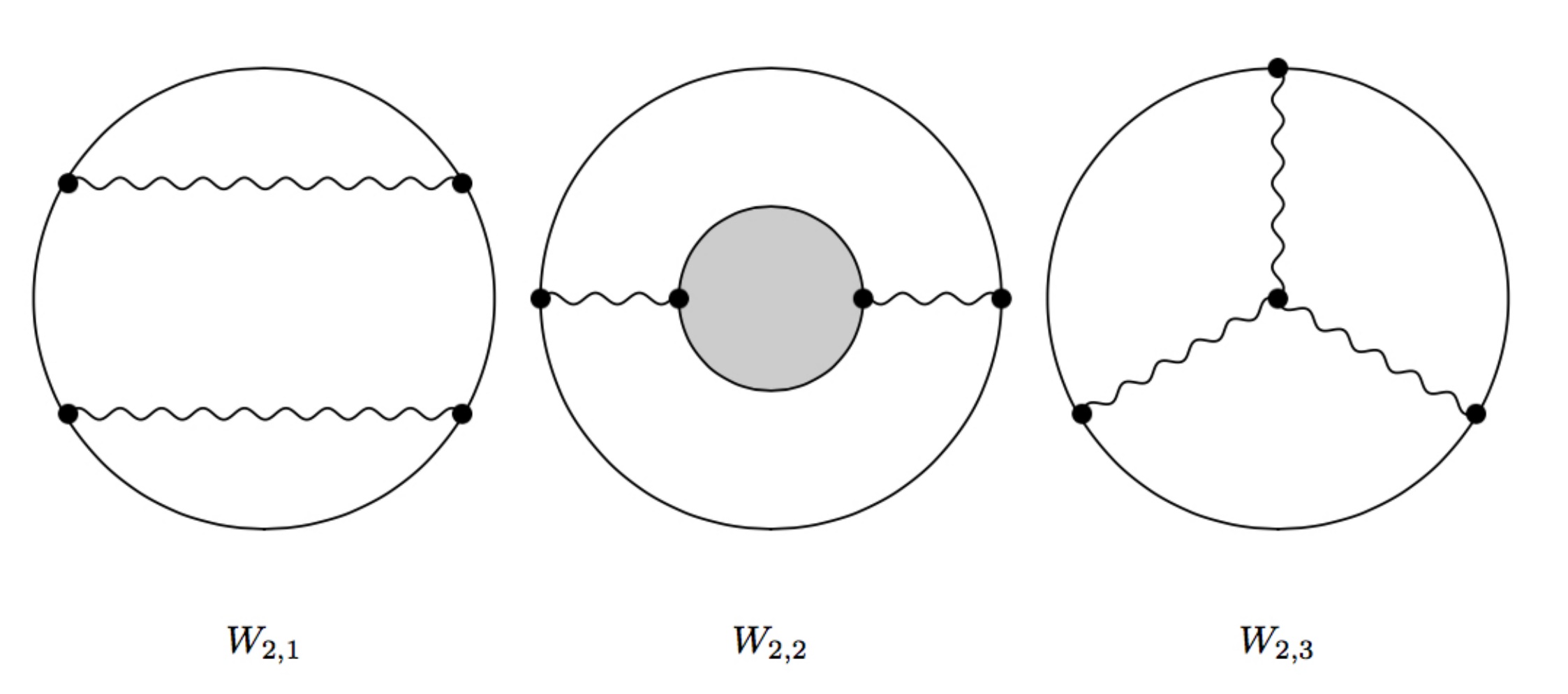}
\caption[] {
Order $\l^2$  contributions to the standard  Wilson loop.
 The middle   diagram  contains  the full self-energy   1-loop  correction
 in  SYM  theory (with   vector,  ghost, scalar  and fermion fields  in the loop). 
For  the Wilson-Maldacena loop there are additional   diagrams   with 
 scalar  propagators instead of some of the vector ones. 
 
\label{fig:two-loops}}
\end{figure}

At order $\lambda^{2}$ there are three types  of  planar contributions
  to  the   Wilson loop in \rf{313}  shown in Fig. \ref{fig:two-loops}  that we shall denote as 
  \be 
  \la{313} \WZ_2 =  \WZ_{2,1}     + \WZ_{2,2}  + \WZ_{2,3}\ .   
  \ee
In the WML   case  it was  found  in  \cite{Erickson:2000af}  that 
 the ladder diagram  contribution 
$\WW_{2,1}$ is finite.
While the  self-energy  part 
$\WW_{2,2}$  and the internal-vertex  part  $\WW_{2,3}$ are  separately   logarithmically 
divergent (all power divergences cancel out in WML case),  their sum is   finite;  moreover,   
the  finite part  also  vanishes in 4 dimensions (in Feynman gauge)
\be\la{2212}
 \WW_{2,2} +\WW_{2,3}=0\ . \ee
In  the WL case, using dimensional regularization to discard power divergences, 
we find  that  the ladder diagram $\W_{2,1}$ in Fig. \ref{fig:two-loops} 
has a logarithmic singularity (i.e. a pole in $\ve=2-\om$). 
 The same is true for both the self-energy diagram $\W_{2,2}$ and 
the internal-vertex diagram $\W_{2,3}$. However, 
 their   sum  in  \rf{313}
turns   out to be  finite 
  (in agreement with the general expectation for a conformal  WL operator 
    in a theory where  the gauge coupling is  not  running).\foot{If one  uses  power   UV cutoff  $a\to 0$ the  remaining  power divergences  universally factorize as an   exponential  factor
$\exp(-k\,{L\ov a})$ where $L$ is the loop length.
This  can be  interpreted  as a mass renormalization of a test particle moving along the loop.
}

Let us  now discuss each  of these contributions in turn.

\begin{figure}[t]  
\centering
\includegraphics[scale=0.35]{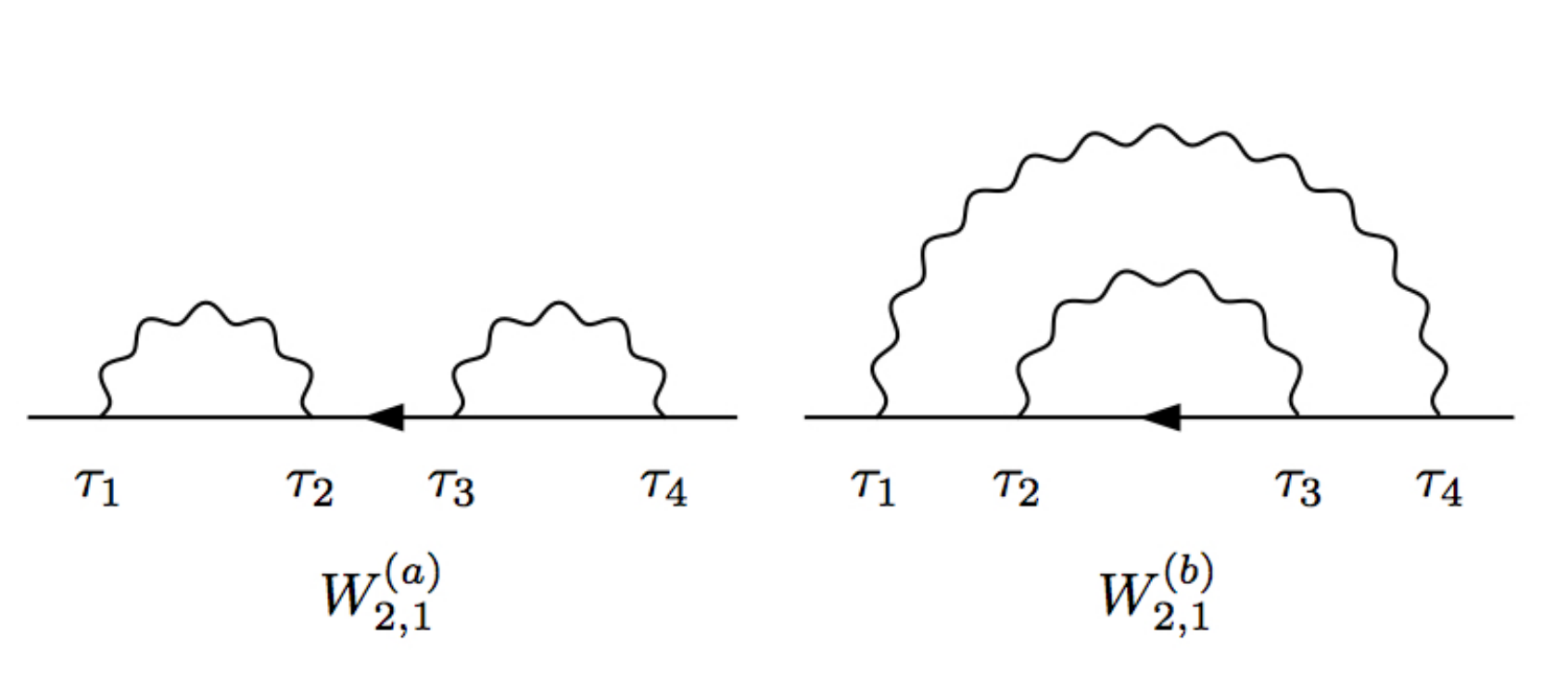}
\caption[] {
Two  of planar diagrams of ladder type $W_{2,1}=W^{(a)}_{2,1} +W^{(b)}_{2,1}   $    with  path-ordered 
four points $\tau_{1}, \dots, \tau_{4}$ in the WL  ($\z=0$) case. For  general $\z$  one needs also to add  similar diagrams  
with scalar propagators. 
\label{fig:rainbow-ladder}}
\end{figure}

\subsubsection{Ladder  contribution } 

The planar ladder diagram $W_{2,1}$ in Fig. \ref{fig:two-loops}  arises from the quartic  term in the expansion of the Wilson loop   operator \rf{0}.
It is convenient to split the integration region 
into $4!$ ordered domains, {\em i.e.} $\tau_{1}>\tau_{2}>\tau_{3}>\tau_{4}$ and similar ones. Before the 
Wick contractions, all these are equivalent and cancel the $4!$ factor  from the 
expansion of the exponential.   
There are two  different planar Wick contractions    shown in Fig. \ref{fig:rainbow-ladder}. 

In the WML   case  the  expression  for the first one  is  \cite{Erickson:2000af}\foot{Here $x^{(i)}=x(\tau_i)$  and $d^{4}\bm{\tau}  \equiv d\tau_{1}d\tau_{2}d\tau_{3}d\tau_4$.}
\be
\la{3.17}
\WW_{\wa} = \frac{\big[\Gamma(\omega-1)\big]^{2}}{64\,\pi^{2\omega}}\,\oint_{\tau_{1}>\tau_{2}>\tau_{3}>\tau_{4}}
d^{4}\bm\tau\,\frac{(|\dot x^{(1)}|\,|\dot x^{(2)}|-\dot x^{(1)}\cdot \dot x^{(2)})
(|\dot x^{(3)}|\,|\dot x^{(4)}|-\dot x^{(3)}\cdot \dot x^{(4)})}
{(|x^{(1)}-x^{(2)}|^{2}\,|x^{(3)}-x^{(4)}|^{2})^{\omega-1}}.
\ee
The second diagram has a similar expression with $(1,2,3,4)\to (1,4,2,3)$.
In the WML  case  these  two  contributions  are equal  and finite. Setting   $\om=2$ 
we find that  the integrand in (\ref{3.17}) in the circle case 
is constant  as in \rf{399}.
As a result, 
\be\la{3.19} 
\WW_{2,1} = \WW_\wa +\WW_{\wb} 
= 2\times \frac{1}{64\,\pi^{4}}\,\frac{(2\pi)^{4}}{4!}\,\Big(\frac{1}{2}\Big)^{2}
 = \frac{1}{192} \ . 
\ee
This  already reproduces  the  coefficient of the $\l^2$ term in \rf{1.1}
(consistently  with  the  vanishing \rf{2212}   of the rest of the   contributions 
\cite{Erickson:2000af}).  

The corresponding   expression in the  WL case   is found 
 by dropping the scalar  field exchanges, \ie\,  the $|\dot x |$ terms in
 the numerator of  \rf{3.17}.  Then for the circle we get  
\begin{align}
\la{3.20}
\W_{\wa} &= \frac{[\Gamma(\omega-1)]^{2}}{64\,\pi^{2\omega}}\,
\int_{\tau_{1}>\tau_{2}>\tau_{3}>\tau_{4}}d^{4}\bm\tau\,
\frac{\cos \tau_{12}\,\cos\tau_{34}}{(4\,\sin^{2}\frac{\tau_{12}}{2}\,4\,\sin^{2}\frac{\tau_{34}}{2})
^{\omega-1}}\ , \notag \\
\W_{\wb} &= \frac{[\Gamma(\omega-1)]^{2}}{64\,\pi^{2\omega}}\,
\int_{\tau_{1}>\tau_{2}>\tau_{3}>\tau_{4}}d^{4}\bm\tau\,
\frac{\cos \tau_{14}\,\cos\tau_{23}}{(4\,\sin^{2}\frac{\tau_{14}}{2}\,4\,\sin^{2}\frac{\tau_{23}}{2})
^{\omega-1}}\ .
\end{align}
The computation of  these integrals  is discussed in Appendix ~\ref{B}. Setting $\om=2-\ve$  we get 
\begin{align}
\la{3.21}
\W_{\wa} &=\te  \frac{[\Gamma(1-\varepsilon)]^{2}}{64\,\pi^{2\,(2-\varepsilon)}}\,
\Big[\frac{\pi^{2}}{\varepsilon}+3\,\pi^{2}+\frac{\pi^{4}}{6}+\mc O(\varepsilon)\Big] = 
\frac{1}{64\,\pi^{2}\,\varepsilon}+\frac{1}{384}+\frac{3}{64\,\pi^{2}}+
\frac{\gamma_{\rm E}+\log\pi}{32\,\pi^{2}}
+\mc O(\varepsilon),\notag \\
\W_{\wb} &=\te  \frac{[\Gamma(1-\varepsilon)]^{2}}{64\,\pi^{2\,(2-\varepsilon)}}\,
\Big[\frac{\pi^{2}}{2}+\frac{\pi^{4}}{6}+\mc O(\varepsilon)\Big] = 
\frac{1}{384}+\frac{1}{128\,\pi^{2}}+\mc O(\varepsilon).
\end{align}
The total ladder contribution in the WL  case  is thus
\be
\la{3.22}
\W_{2,1} =\W_{\wa}+ \W_{\wb}
= \frac{1}{64\,\pi^{2}\,\varepsilon}+\frac{1}{192}+\frac{7}{128\,\pi^{2}}
+\frac{\gamma_{\rm E}+\log\pi}{32\,\pi^{2}}+\mc O(\varepsilon).
\ee

\subsubsection{Self-energy contribution} 
\la{sec:self-wilson}

It is   convenient to represent the  contribution $W_{2,2}$  of the self-energy 
diagram  in Fig. \ref{fig:two-loops}  as 
\be
\la{3.23}
\WZ_{2,2} = - \, \frac{[\Gamma(\omega-1)]^2}{8\,\pi^{\omega}(2-\omega)(2\omega-3)}\, \wWZ_{1} \ , 
\ee
where, in the  WML  case,  one has  \cite{Erickson:2000af}
\be
\la{3.24}
 {\wWW_{1}}  = \frac{1}{16\pi^{\omega}}\oint d\tau_{1}d\tau_{2}\,
\frac{|\dot x(\tau_{1})|\,|\dot x(\tau_{2})|-\dot x(\tau_{1})\cdot \dot x(\tau_{2})}{
\big[|x(\tau_{1})-x(\tau_{2})|^{2}\big]^{2\omega-3}}  \ . 
\ee
Again, the  expression 
 in the  WL  case is   obtained   by  simply dropping  the  scalar exchange 
  $|\dot x(\tau_{1})|\,|\dot x(\tau_{2})|$ term in the numerator of (\ref{3.24}):
  \be
\la{3.25}
 {\wwW_1}  = \frac{1}{16\pi^{\omega}}\oint d\tau_{1}d\tau_{2}\,
\frac{-\dot x(\tau_{1})\cdot \dot x(\tau_{2})}{
\big[|x(\tau_{1})-x(\tau_{2})|^{2}\big]^{2\omega-3}}  \ . 
\ee
Altough (\ref{3.25})  is  very similar to $\W_{1}$  in \rf{3.8},  for $\om\not=2$ 
there is a  difference  in the power in the denominator.
\iffa 
We   may write 
\begin{align}
\la{3.21}
\W_{2,2} &= \WW_{2,2}
-\frac{[\Gamma(\omega-1)]^{2}}{8\,\pi^{\omega}(2-\omega)(2\omega-3)}\,(\widetilde \W_{1}-\widetilde \WW_{1}).
\end{align}
To compute this quantity we need the expansion of $\widetilde \W_{1}$ and 
$\widetilde \WW_{1}$ to order  $\varepsilon$ included.
\fi
 Specializing to the circle case  we  find  (using the integral  \rf{3.12})
\begin{align}
 \wWW_{1} &=  
 2^{3-4\omega}\pi^{-\omega}\oint d\tau_{1}d\tau_{2}\,
\Big[\sin^{2}\tfrac{\tau_{12}}{2}\Big]^{4-2\omega} = \frac{2^{3-4\omega}\pi^{-\omega}}{2\pi}\,\mc M(4-2\omega,0,0)  \notag \\
&= 
\frac{1}{8}+\frac{1}{8}\log\pi\, \varepsilon+\OO(\ve^2) \ , \la{3.26}\\
 \wwW_{1} &=  
 -4^{1-2\omega}\pi^{-\omega}\oint d\tau_{1}d\tau_{2}\,
\Big[\sin^{2}\tfrac{\tau_{12}}{2}\Big]^{3-2\omega} 
+2^{3-4\omega}\pi^{-\omega} \oint d\tau_{1}d\tau_{2}\,
\Big[\sin^{2}\tfrac{\tau_{12}}{2}\Big]^{4-2\omega}\notag \\
&= \frac{1}{8}+\frac{1}{8}(2+\log\pi)\,\varepsilon+ \OO(\ve^2). \la{3.27}
\end{align}
Then from  (\ref{3.23})  we get 
\begin{align}
\la{3.28}
\WW_{2,2} &= -\frac{1}{64\,\pi^{2}\,\varepsilon}-\frac{1}{32\,\pi^{2}}-\frac{\gamma_{\rm E}}
{32\,\pi^{2}}-\frac{\log\pi}{32\,\pi^{2}}+\OO(\ve),  \\
\W_{2,2} &= -\frac{1}{64\,\pi^{2}\,\varepsilon}-\frac{1}{16\,\pi^{2}}-\frac{\gamma_{\rm E}}
{32\,\pi^{2}}-\frac{\log\pi}{32\,\pi^{2}}+\OO(\ve) \ . \la{3.29}
\end{align}
Note that  the  difference between the  WL  and WML   self-energy contributions is finite 
\begin{align}
\la{3.30}
\W_{2,2} 
= \WW_{2,2}-\frac{1}{32\,\pi^{2}}  \ .
\end{align}

\subsubsection{Internal-vertex  contribution}  

In the WML   case, the   internal-vertex diagram  contribution  in Fig. 2  has 
the following expression \ci{Erickson:2000af}
\begin{align}
\WW_{2,3} &= -\frac{1}{4}\,\oint d^{3}\bm{\tau}  \, \varepsilon(\tau_{1},
\tau_{2},\tau_{3})\ \Big[|\dot x^{(1)}|\,|\dot x^{(3)}|-\dot x^{(1)}\cdot \dot x^{(3)}\Big]\notag \\
&\ \qquad \qquad \times \dot x^{(2)}\cdot \frac{\partial}{\partial x^{(1)}}\int d^{2\omega}y\,
\Delta(x^{(1)}-y)\,\Delta(x^{(2)}-y)\,\Delta(x^{(3)}-y),\la{3.31}
\end{align}
where  $\Delta(x)$  is the  propagator  \rf{3.7}, $d^{3}\bm{\tau}  \equiv d\tau_{1}d\tau_{2}d\tau_{3}$
and $\varepsilon(\tau_{1},\tau_{2},\tau_{3})$ is 
the totally  antisymmetric path ordering symbol equal to  $1$ if $\tau_{1}>\tau_{2}>\tau_{3}$.
 Using  the Feynman parameter  representation for the   propagators  and specializing to the circle case   \rf{3.31} becomes  
\begin{align}
\la{3.32}
&\WW_{2,3} = \frac{\Gamma(2\omega-2)}{2^{2\omega+5}\,\pi^{2\omega}}\,
\int_{0}^{1} [d^3\bm{\a}]  \  \oint      d^{3}\bm{\tau}      \,\eps(\tau_{1},\tau_{2},\tau_{3})\       \notag\\
&  \qquad \qquad \qquad \qquad \qquad \qquad \times 
\big(1-\cos\tau_{13}\big)\ \frac{\alpha\,(1-\alpha)\,\sin\tau_{12}
+\alpha\,\gamma\,\sin\tau_{23}}{Q^{2\,\omega-2}}, \\
&\qquad \qquad  [d^3\bm{\a}] \equiv d\alpha\,d\beta\,d\gamma\,(\alpha\beta\gamma)^{\omega-2}\,
\delta(1-\alpha-\beta-\gamma)  \ , \\
&\qquad \qquad Q \equiv  \alpha\,\beta\,(1-\cos\tau_{12})+\beta\,\gamma\,(1-\cos\tau_{23})
+\gamma\,\alpha\,(1-\cos\tau_{13}) \ . 
\la{3.33} \end{align}
The corresponding  WL  expression is  found by 
omitting the scalar coupling term $|\dot x^{(1)}|\,|\dot x^{(3)}|$, \ie\space 
by replacing the factor $(1-\cos\tau_{13}) $  by $( -\cos\tau_{13}) $.
We can then   represent the WL   contribution as 
\begin{align} 
\la{3.34}
\W_{2,3} &= \WW_{2,3}  -\frac{\Gamma(2\omega-2)}{2^{2\omega+5}\pi^{2\omega}}\,
J(\omega),
\\
\la{3731}
J(\omega) &\equiv 
\int_{0}^{1}  [d^3\bm{\a}]   
 \oint d^{3}\bm{\tau}\,\eps(\tau_{1},\tau_{2},\tau_{3})\ \frac{\alpha\,(1-\alpha)\,\sin\tau_{12}
+\alpha\,\gamma\,\sin\tau_{23}}{Q^{2\,\omega-2}} \ . 
\end{align}
In the WML   case one finds that \rf{3.32}  is related to    $\WW_{2,2}$ \cite{Erickson:2000af}  
\be
\la{3.36}
\WW_{2,3} =- \WW_{2,2} + \mc O(\varepsilon) \ , 
\ee
where $\WW_{2,2}$    was given in \rf{3.28}. Thus 
to compute  $\W_{2,3}$ it remains to  determine  $J(\omega)$.
Let   us  first   use that 
\begin{align}
& \oint d^{3}\bm{\tau}\,\varepsilon(\tau_{1},\tau_{2},\tau_{3}) \ F(\tau_{1},\tau_{2},\tau_{3}) =\oint_{\tau_{1}>\tau_{2}>\tau_{3}}d^{3}\bm{\tau}\,\Big[
 F(\tau_{1}, \tau_{2}, \tau_{3})
-F(\tau_{1}, \tau_{3}, \tau_{2}) \notag \\
&\qquad
+F(\tau_{2}, \tau_{3}, \tau_{1})
-F(\tau_{2}, \tau_{1}, \tau_{3}) +F(\tau_{3}, \tau_{1}, \tau_{2})
-F(\tau_{3}, \tau_{2}, \tau_{1})\Big], \la{3355}
\end{align}
and relabel the Feynman parameters in each term. Then  $J(\omega)$  takes    a  more symmetric form 
\begin{align}
\la{3.38}
J(\omega) &= 
8\,\int_{0}^{1}  [d^3\bm{\a}] 
\oint_{\tau_{1}>\tau_{2}>\tau_{3}} d^{3}\bm{\tau}\,\frac{(\alpha\,\beta+\beta\,\gamma
+\gamma\,\alpha)\,\sin \frac{\tau_{12}}{2}\,\sin \frac{\tau_{13}}{2}\,\sin \frac{\tau_{23}}{2}}
{Q^{2\,\omega-2}} \ . 
\end{align}
Using the double Mellin-Barnes representation (see,  for instance, \cite{Jantzen:2012cb})
\be\la{3377}
\frac{1}{(A+B+C)^{\s}} = \frac{1}{(2\,\pi\,i)^{2}}\frac{1}{\Gamma(\s)}\,
\int_{-i\,\infty}^{+i\,\infty}du\,dv\,\frac{B^{u}\,C^{v}}{A^{\s+u+v}}\,
\Gamma(\s+u+v)\,\Gamma(-u)\,\Gamma(-v),
\ee
we can further rewrite (\ref{3.38}) as
\begin{align}
&J(\omega) = \frac{8}{(2\pi i)^{2}\,2^{2\omega-2} \Gamma(2\omega-2)}
\oint_{\tau_{1}>\tau_{2}>\tau_{3}} d^{3}\bm{\tau}
\int du dv 
 \int_{0}^{1}d\alpha\, d\beta\, d\gamma\,  (\alpha\beta\gamma)^{\omega-2}(\alpha\beta
+\beta\gamma+\gamma\alpha)\la{3.40} \\
&\times \Gamma(2\omega-2+u+v)\Gamma(-u)\Gamma(-v)\,
\frac{(\beta\gamma\sin^{2}\frac{\tau_{23}}{2})^{u}
\,(\alpha\beta\,\sin^{2}\frac{\tau_{12}}{2})^{v}}
{(\gamma\alpha\,\sin^{2}\frac{\tau_{13}}{2})^{2\omega-2+u+v}} 
\sin \tfrac{\tau_{12}}{2}\,\sin \tfrac{\tau_{13}}{2}\,\sin \tfrac{\tau_{23}}{2}\ .\no 
\end{align}
Integrating over  $\alpha$,$\beta$,$\gamma$ 
  using 
  the relation
\be\la{3.41}
\int_{0}^{1}\prod_{i=1}^{N} d\a_{i}\,\a_{i}^{\nu_{i}-1}\,\delta(1-\sum_{i} \a_{i}) = 
\frac{\Gamma(\nu_{1})\cdots\Gamma(\nu_{N})}{\Gamma(\nu_{1}+\cdots+ \nu_{N})} \ , 
\ee
 gives the following representation for $J$
\begin{align}
J(\omega) =& -\frac{1}{\pi^{2}\,2^{2\omega-3}}\frac{1}{\Gamma(2\omega-2)\Gamma(3-\omega)}
\int_{-i\infty}^{+i\infty} du \int_{-i\infty}^{+i\infty}dv\, \  X(u,v)\ T(u,v)\  ,   \la{3.42} \\  &\
X(u, v) \equiv  \Big(\frac{1}{u+v+\omega-1}-\frac{1}{u+\omega-1}-\frac{1}{v+\omega-1}\Big)
\la{3.43}  \\
& \qquad  \times \Gamma(2\omega-2+u+v)\Gamma(-u)\Gamma(-v)\,
\Gamma(2-u-\omega)\,\Gamma(2-v-\omega)\,
\Gamma(u+v+\omega)\ , \ \  \notag \\
&T(u,v) \equiv\oint_{\tau_{1}>\tau_{2}>\tau_{3}} d^{3}\bm{\tau}\,
\frac{(\sin^{2}\frac{\tau_{23}}{2})^{u+1/2}\,(\sin^{2}\frac{\tau_{12}}{2})^{v+1/2}} 
{(\sin^{2}\frac{\tau_{13}}{2})^{2\omega-2+u+v-1/2}}\ .\la{344}
\end{align}
A remarkable feature of (\ref{3.42}), familiar in computations of  similar  integrals, 
 is that the integrand
 is symmetric in the three $\tau_i$ variables  as one can  show using a 
  suitable linear change of  the Mellin-Barnes integration parameters $u,v$.\footnote{
For instance, the exchange of $\tau_{1}$ and $\tau_{3}$ is compensated by redefining 
$(u,v)\to (u',v')$ with 
$ u+\frac{1}{2} = -(2\omega-2+u'+v'-1/2),\ \
 -(2\omega-2+u+v-1/2) = u'+1/2, $ that is 
$
u=2-u'-v'-2\omega, \ v=v'.
$
This change of variables leaves invariant the other part  $T(u,v)$ of the integrand: it 
takes the  same form when written in terms of $u',v'$.}
As a result, we may  effectively replace $T(u,v)$   by   $1\ov 3!$ of the  integrals 
along the full circle: 
\begin{align}
 T(u,v) \to  {1 \ov 3!}  \oint_{0}^{2\pi}d^{3}\bm{\tau}\,
\frac{(\sin^{2}\frac{\tau_{23}}{2})^{u+1/2}\,(\sin^{2}\frac{\tau_{12}}{2})^{v+1/2}}
{(\sin^{2}\frac{\tau_{13}}{2})^{2\omega-2+u+v-1/2}}. \la{355}
\end{align}
Using  again  the master integral (\ref{3.12}), we find  the  following expression for $J(\omega)$ as a double integral
\begin{align}
J(\omega) &= -\frac{8\,\pi^{3/2}}{3!\,\pi^{2}\,2^{2\omega-3}}\frac{1}{\Gamma(2\omega-2)
\Gamma(3-\omega)}
\int_{-i\infty}^{+i\infty} du \int_{-i\infty}^{+i\infty}dv\,  X(u,v)  \notag \\
&\qquad \qquad \times 
\frac{\Gamma (u+1) \,\Gamma (v+1)\, \Gamma \Big(\frac{9}{2}-2 \omega \Big)\, \Gamma (-u-v-2
   \omega +3)}{\Gamma (u+v+2) \,\Gamma (-u-2 \omega +4) \,\Gamma (-v-2 \omega +4)}. \la{366}
\end{align}
Writing all factors  in  $X(u,v)$  in \rf{3.43}   in terms of $\Gamma$-functions
we end up with  
\iffa \footnote{The  coefficient here comes from 
$-\frac{8\,\pi^{3/2}}
{3!\,\pi^{2}\,2^{2\omega-3}}(2\,\pi\,i)^{2} = \frac{\pi^{3/2}}{3\times 2^{2\omega-7}}.
$
}\fi
\begin{align}
&J(\omega) = 
\frac{\pi^{3/2}}{3\times 2^{2\omega-7}}\,
\frac{1}{\Gamma(2\omega-2)
\Gamma(3-\omega)}
\int_{-i\infty}^{+i\infty} \frac{du}{2\,\pi\,i} \int_{-i\infty}^{+i\infty} \frac{dv}{2\,\pi\,i}\  R(u,v) \la{3.47} \ , \\
& R(u,v) =  \Gamma(2\omega-2+u+v)\Gamma(-u)\Gamma(-v) \,\Big[
 \Gamma(1-u-\omega)\,\Gamma(2-v-\omega)\,\Gamma(u+v+\omega)\,\notag \\
& +\Gamma(2-u-\omega)\,\Gamma(1-v-\omega)\,\Gamma(u+v+\omega) +\Gamma(2-u-\omega)\,\Gamma(2-v-\omega)\,\Gamma(u+v+\omega-1)\Big]\,\notag \\
&\qquad \qquad \times 
\frac{\Gamma (u+1) \Gamma (v+1) \Gamma \Big(\frac{9}{2}-2 \omega \Big) \Gamma (-u-v-2
   \omega +3)}{\Gamma (u+v+2) \Gamma (-u-2 \omega +4) \Gamma (-v-2 \omega +4)}.
\end{align}
This integral can be computed using  the algorithms described in \cite{Czakon:2005rk} and by 
repeated application of   Barnes first and second lemmas \cite{bailey1935generalized}. 
The result  expanded in $\ve = 2 - \om \to 0$ 
is 
\be\la{3.49}
J(2-\varepsilon) = \frac{8\,\pi^{2}}{\varepsilon}-8\,\pi^{2}\,(2\,\log 2-3)+ \mc O(\varepsilon).
\ee
Using this in  (\ref{3.34})  gives 
\be
\la{3.50}
\W_{2,3} = \WW_{2,3}-\frac{1}{64\,\pi^{2}\,\varepsilon}
-\frac{1}{64\,\pi^{2}}-\frac{\gamma_{\rm E}+\log\pi}{32\pi^{2}}+ \mc O(\varepsilon).
\ee

\subsubsection{Total  contribution to standard Wilson loop}

From  \rf{3.30}   and \rf{3.50}   we get  
\be  
\W_{2,2}  + \W_{2,3} = -\frac{1}{64\,\pi^{2}\,\varepsilon}
-\frac{3}{64\,\pi^{2}}-\frac{\gamma_{\rm E}+\log\pi}{32\pi^{2}}+ \mc O(\varepsilon).
\la{3667}\ ,  \ee
\ie\space in contrast to the WML case  \rf{2212},\rf{3.36}  the sum of the self-energy and   internal vertex    diagrams 
is no longer  zero and is  logarithmically divergent.
The divergence   is cancelled once we add the ladder  contribution in \rf{3.22}.
Thus 
the  total  contribution to the WL expectation value 
at order $\l^2$    found  from \rf{3.22},\rf{3667}  is   finite 
\begin{align}\la{350}
&\W_{2} = \W_{2,1}+\W_{2,2}+\W_{2,3} = \frac{1}{192}+\frac{1}{128\,\pi^{2}} \ , \qquad \qquad 
\W_2 =\WW_2 + \frac{1}{128\,\pi^{2}} \ .  \end{align}
Thus, using \rf{3.3},\rf{3466}, we get the final result for the expectation value of the ordinary Wilson loop  
\begin{align} &\la{3511}
 \W = 1 +  {1\ov 8} \l  +  \Big(\frac{1}{192}+\frac{1}{128\,\pi^{2}}\Big)\,\l^2  +  \OO(\l^3) \ . 
\end{align}
We conclude that     the   weak-coupling 
expectation values for  the  circular 
WML and WL start to differ from  order $\l^2$.

\subsection{Generalization to   any  $\zeta$  }

Let   us now generalize   the   above results  for the leading and subleading 
term in the weak-coupling   expansion  \rf{3.3} of the circular Wilson loop
to the  case of the  generalized WL, i.e. to any value of the parameter $\z$ in \rf{0}. 
The computation  follows the same lines as  above. 

At leading order in $\l$ we find the same result as in the  circular  WML \rf{3.6} 
 and WL  \rf{3466}  cases, 
\ie,     after subtracting the linear divergence,   the quantity $W_1$ in \rf{3.3} 
  has the universal (independent on $\z$) value
\be  \la{3.53} 
W^{(\z)}_1 = {  1\ov 8}  \ . \ee
Explicitly, using again dimensional regularization, we find  as in  \rf{3.4},\rf{3.11},\rf{3331}
 \begin{align}
 \WZ_{1} &=\frac{ \G(\om-1)   }{16\pi^{\om }}\oint d\tau_{1}d\tau_{2}\,
\frac{\zeta^{2}-\dot x(\tau_{1})\cdot \dot x(\tau_{2})}{
|x(\tau_{1})-x(\tau_{2})|^{2\omega-2}}\no \\ 
&= \frac{1}{8} 
-\frac{(1-\zeta^{2})\,  \G(\om-1)}{16\pi^{\omega}}\oint {d\tau_{1}d\tau_{2}\ov \big(4\,
\sin^{2} \tfrac{\tau_{12}}{2}\big)^{\omega-1} }
= \frac{1}{8}+  \frac{1}{8}(1-\z^2) \ve   +    \OO(\ve^2)   \la{3.54}
 \end{align}
where we  set $\omega=2-\ve$ and  retained a term of order $ \ve$ as this will contribute to the
 final result at order $\lambda^2$ in our 
dimensional regularization scheme upon replacing the 
bare with renormalized coupling. To order $\lambda$, however, one can safely 
remove this term yielding (\ref{3.53}). 

Turning to $\l^2$ order,  the ladder  diagram contributions in Fig. \ref{fig:two-loops} generalizing the $\z=0$   expressions \rf{3.20} are 
\begin{align}
W^{(\z)}_{\wa} &= \frac{[\Gamma(\omega-1)]^{2}}{64\,\pi^{2\omega}}\,
\int_{\tau_{1}>\tau_{2}>\tau_{3}>\tau_{4}}d^{4}\bm\tau\,
\frac{(\zeta^{2}-\cos \tau_{12})\,(\zeta^{2}-\cos\tau_{34})}{(4\,\sin^{2}\frac{\tau_{12}}{2}\ 4\,\sin^{2}\frac{\tau_{34}}{2})
^{\omega-1}}, \notag \\
W^{(\z)}_{\wb} &= \frac{[\Gamma(\omega-1)]^{2}}{64\,\pi^{2\omega}}\,
\int_{\tau_{1}>\tau_{2}>\tau_{3}>\tau_{4}}d^{4}\bm\tau\,
\frac{(\zeta^{2}-\cos \tau_{14})\,(\zeta^{2}-\cos\tau_{23})}{(4\,\sin^{2}\frac{\tau_{14}}{2}\ 4\,\sin^{2}\frac{\tau_{23}}{2})
^{\omega-1}}.\la{3543}
\end{align}
The result of their rather involved  computation generalizing  \rf{3.21}    is  (see Appendix \ref{B}) 
\begin{align}
W^{(\z)}_{\wa} &= \frac{[\Gamma(1-\varepsilon)]^{2}}{64\,\pi^{2\,(2-\varepsilon)}}\,\Big[
\frac{\pi^{2}\,(1-\zeta^{2})}{\varepsilon}+\pi^{2}\,(1-\z^2) ( 3 - \z^2) 
+\frac{\pi^{4}}{6}
+\mc O(\varepsilon)
\Big], \notag \\
W^{(\z)}_{\wb} &= \frac{[\Gamma(1-\varepsilon)]^{2}}{64\,\pi^{2\,(2-\varepsilon)}}\,\Big[
\frac{\pi^{2}}{2}\,(1-\zeta^{2})^{2}+\frac{\pi^{4}}{6}
+\mc O(\varepsilon)\Big],\la{3556} 
\end{align}
with the sum  being 
\be
W_{2,1}^{(\z)} = W^{(\z)}_{\wa}  + W^{(\z)}_{\wb} =\frac{1}{192}+(1-\zeta^{2})\,\Big[
\frac{1}{64\,\pi^{2}\,\varepsilon}+\frac{1}{128\,\pi^{2}}\,(7-3\,\zeta^{2})
+\frac{\log\pi+\gamma_{\rm E}}{32\,\pi^{2}} \Big] + \mc O(\varepsilon) \ . 
\la{8899}
\ee
For the self-energy  contribution  in Fig. \ref{fig:two-loops}   we find  the expression \rf{3.23} 
where now 
\begin{align}
{\widetilde W}^{(\z)}_{1} &= \frac{1}{16\pi^{\omega}}\oint d\tau_{1}d\tau_{2}\ 
\frac{\zeta^{2}\,|\dot x(\tau_{1})|\,|\dot x(\tau_{2})|-\dot x(\tau_{1})\cdot \dot x(\tau_{2})}{
\big[|x(\tau_{1})-x(\tau_{2})|^{2}\big]^{2\omega-3}} \notag \\
&= \zeta^{2}\,\widetilde W_{1}^{(1)}+(1-\zeta^{2})\,\widetilde W_{1}^{(0)}
= \frac{1}{8}+\frac{1}{8}\,\Big[2\,(1-\zeta^{2})+\log\pi\Big]+  \mc O(\varepsilon), \la{3365}
\end{align}
with ${\widetilde W}_{1}^{(1)}$ and ${\widetilde W}_{1}^{(0)}$    given by \rf{3.24},\rf{3.26}  and \rf{3.25},\rf{3.27}. 
Substituting this into  (\ref{3.23}),  we get 
\be\la{3577}
\WZ_{2,2} = \zeta^{2}\,W_{2,2}^{(1)}+(1-\zeta^{2})\,\Big[
-\frac{1}{64\,\pi^{2}\,\varepsilon}-\frac{1}{16\,\pi^{2}}-\frac{\gamma_{E}+\log\pi}{32\,\pi^{2}}
\Big]+\mc O(\varepsilon) \ , 
\ee
where $W_{2,2}^{(1)} $   is given by \rf{3.28}. 

The  internal-vertex diagram contribution   in Fig. \ref{fig:two-loops} 
generalizing \rf{3.34} is 
\be\la{3588}
\WZ_{2,3} = W_{2,3}^{(1)}-(1-\zeta^{2})\,\frac{\Gamma(2\omega-2)}{2^{2\omega+5}
\pi^{2\omega}}\,J(\omega) \ , 
\ee
where $J$  is given by \rf{3731},\rf{3.49}
and $W_{2,3}^{(1)}  $ is given by \rf{3.36},\rf{3.28}, i.e. 
\be\la{3599} 
\WZ_{2,3} =-\WW_{2,2}+(1-\zeta^{2})\,\Big[-\frac{1}{64\,\pi^{2}\,\varepsilon}
-\frac{1}{64\,\pi^{2}}-\frac{\gamma_{\rm E}+\log\pi}{32\pi^{2}}\Big] +  \mc O(\ve)\ .
\ee
Summing  up the  separate   contributions   given in     \rf{8899},\rf{3577}   and  \rf{3599}         
we find that the  ${1\ov \ve} \sim \log a $    logarithmic divergences   cancel out, and  we get the finite expression 
\be
\la{3.61}
\WZ_{2} = \WZ_{2,1}+\WZ_{2,2}+\WZ_{2,3} = \frac{1}{192}
+\frac{1}{128\pi^{2}}\,(1-\zeta^{2})\,(1-3\zeta^{2}) \ .
\ee
The final result for the Wilson loop expectation value to order $\lambda^2$  that  follows from \rf{3.54} and \rf{3.61} 
is then 
\begin{equation}
\la{2633}
\lan \WZ \ran = 1+ \lambda \Big(\frac{1}{8}-\frac{1}{8}\zeta^2\ve \Big)
+\lambda^2 \left[ \frac{1}{192}
+\frac{1}{128\pi^{2}}\,(1-\zeta^{2})\,(1-3\zeta^{2}) \right]  + \OO(\l^3)\ . 
\end{equation}
Here it is  important to retain the order $ \z^2\ve $ part in the 1-loop term in \rf{3.54}:  
 despite the cancellation of all  $1\ov \ve$ terms  to this order, 
$\z$  in the  order $\l$ term is   a bare coupling  that contains  poles that may  
effectively contribute  at higher orders.  

Despite  $\l$   not running  in $d=4$  the presence of the linear in $\z$ term in the beta-function 
\rf{111}   implies   that the present case is best treated as a 2-coupling $g_i=(\l, \z)$  theory.  In    general, if $d= 4- 2 \ve$  
and we have a set of near-marginal couplings $g_i$  with 
mass dimensions  $ u_i\ve$   the bare couplings may be expressed in terms of the dimensionless  renormalized  couplings  $g_i$ as 
\begin{align} \la{267}
&  g_i{}_{\bb} = \mu^{u_i\ve}\Big[g_i+\frac{1}{\ve} K_i(g) + \OO\big({1\ov \ve^2}\big)\Big]\,,\qquad \qquad \mu { d  g_i{}_\bb \ov d \mu}=0, 
\\
& \la{268} 
  \beta_i(g) = \mu { d g_i \ov d \mu}
= -  \ve   u_i g_i     - u_i K_i   +  \sum_j  u_j g_j {\del\ov \del g_j}   K_i   \ . 
\end{align}
In the present case we may choose  dimensions so that the gauge field  and scalars $\P_m$   in the bare   SYM action 
$ {N\ov \l_\bb} \int d^{d} x ( F^2+ D\P  D\P + ...) $  have dimension 1  so that $\l_b$   
has dimension $2\ve $, 
\ie\space  $\l_\bb = \mu^{2 \ve} \l$,  or $u_\l=  2$  (and of course $K_{\lambda}=0$). 
As the Wilson line  integrand   in \rf{0}   should have dimension 1, that means $\z_b$ should  have  dimension zero, \ie \space
 $u_\z=0$.\foot{
 This is   natural   as the dimension  of the  Wilson line  integral is not changed. Note that the same  is true if one 
 redefines  the SYM  fields   by a power of gauge coupling $g$: then dimension of $\P$ is  canonical ${d-2\ov 2}= 1-\ve$  
 but $g\, \P$   that then enters the Wilson loop  \rf{0} still has  dimension 1.}
  Then   from \rf{267},\rf{268} we learn that   (using \rf{111})
\be \la{269}
\z_\bb = \z + \frac{1}{\ve} K_\z  + \OO\big({1\ov \ve^2}\big) \ , \ \ \ \ \ \ \ \
\beta_\z= u_\l  \l {\del\ov \del \l}   K_\z\ , \qquad \qquad    K_\z = {1\ov 2} \beta_\z = \frac{\lambda}{16\pi^2}\z (\z^2-1)  \ . 
\ee
The coupling $\z$    in \rf{2633}  should actually be the bare coupling;
replacing it  with the renormalized coupling according to \rf{269} 
and  then  sending $\ve \to 0$   we  find   the expression in \rf{1}, i.e. 
\begin{align} &\la{3500}
\lan \WZ\ran = 1 +  {1\ov 8} \l  +   \left[\frac{1}{192}
+\frac{1}{128\pi^{2}}\,(1-\zeta^{2})^2\,\right]\l^2  +  \OO(\l^3) \ .  
\end{align}
As we shall discuss in Appendix \ref{ab3},   there is  an alternative regularization procedure in which the  full 2-loop 
expression  in \rf{3500}  comes just from the type (b) ladder diagram  contribution in \rf{3556} and thus 
the  use of  the evanescent  1-loop term in \rf{2633} is not required.

\iffa 
Note  that the  cancellation of logarithmic   divergences  at order $\l^2$ does not contradict the 
non-trivial renormalization  \rf{111} of $\z$. In general,  renormalizability of $\WZ$  implies that 
  $\lan \WZ\ran \equiv F[\l, \z(a), \log a] =  F[\l, \z(\mu), \log \mu]$
order by order in $\l$  and thus   the finiteness of $\l^2$ term is due to the absence of $\z$ dependence at order $\l$
in \rf{3500}.  The $\z$-dependence of the $\l^2$   term in \rf{3500}   implies the presence of logarithmic UV divergences at  the 
 $\l^3$  and higher orders. 
\fi

\section{Relation  to correlators of  scalar operators on the  Wilson loop}\label{sec4}

The  $\z$-dependence of the generalized WL \rf{0}   can be  viewed as being to due to multiple insertions 
 of  the scalar operators  on the loop.  It is of interest  to relate the expression \rf{3500}  to  what is known about 2-point functions 
 of (scalar) operators  on the line or circle  (see \ci{Polyakov:2000ti,Drukker:2006xg,Alday:2007he,Sakaguchi:2007ba,Drukker:2011za,Correa:2012at,Cooke:2017qgm,Giombi:2017cqn}).
 Let us  choose   the scalar  coupling  in \rf{0}  to be along 6-th direction, \ie\space 
  $\P_m \theta^m =\P_6$ and  denote the remaining 5  scalars  not coupled  directly  to the loop as   $\P_a$  ($a=1,...,5$). Let  us also   choose the contour to be straight line  $x^\m= (\tau, 0, 0, 0)$ along the Euclidean time direction
$x^0 =t$    so that the exponent  in \rf{0}   is simply $ \int dt ( i A_t + \z  \P_6) $.  
For $\z=1$ or $\z=0$   when the loop  preserves the conformal symmetry 
 the   2- (and higher)  point functions  of   conformal operators  inserted  along the line can be interpreted 
 as correlators in  an  effective  (defect) 1d CFT. For example, for $\z=1$
\begin{equation}\la{7}
\llangle   O(t_1)  O(t_2)  \rrangle_{\rm line}  \equiv 
 \langle {\rm Tr}\,{\cal P} \big[O(x_1)O(x_2)\ e^{\int dt(iA_t+ \Phi_6)}\big]\rangle = \frac{C}{|t_{12}|^{2\Delta}}\  
\ . \end{equation}
Here  in $ \langle {\rm Tr} ... \rangle$    the operator $O(x)$  is a gauge-theory  operator in the adjoint representation 
restricted to the line (with exponential factors appearing between and after $O(x_n(t_n))$ according to path ordering to preserve gauge invariance). 
We also   use that in the WML  case for a straight line  the normalization factor is trivial, 
\ie\space  $\lla  1 \rra =1$. 
Similar  relation  can be written for a circular   loop using   the map $t \to \tan{\tau\ov 2}$
\begin{equation}
 \llangle O(\tau_1) O(\tau_2)\rrangle_{\rm circle}=
\frac{C}{|2\sin\frac{\tau_{12}}{2} |^{2\Delta}}\,.
\label{77}
\end{equation}
Here the  gauge-theory expectation value  is to be normalized with  the non-trivial   circle  WML  factor \rf{1.1} 
so that once again $\llangle  1 \rrangle =1$.  In the $\z=0$ case one is to use \rf{3511} as the  corresponding 
normalization factor.
In what follows $\llangle ...\rrangle$   will refer to the expectation value in the effective CFT on the circle.

The simplest example  is the insertion of the ``orthogonal'' scalars $\P_a$   into  the WML \rf{7} 
in which case the dimension is   protected,  $\Delta=1$,  
while    the norm   is related to the Bremsstrahlung function $B(\l)$ \ci{Correa:2012at}\footnote{Let us  recall 
that the  leading tree level value of the 2-point
coefficient $C=\frac{\lambda}{8\,\pi^{2}}+\dots$  (with $\l\equiv  g^2 N$) is found by taking into the account 
that the adjoint  scalar field is $\Phi =  \Phi^r t^r$  with propagator 
 $\langle\Phi^{r}(x)\Phi^{r'}(0)\rangle = \frac{g^{2}\,\delta^{rr'}}{4\pi^{2}\,x^{2}}$ 
($r=1, ..., N^2-1$ is the $SU(N)$ algebra index)  where  
the generators   satisfy  $\text{Tr}(t_{r}t_{r'}) = \frac{1}{2}\,\delta_{rr'}$, \ $t_r t_r =\ha N\,  {\rm I}$. The trace 
 $\delta^{rr'} \delta_{rr'} = N^2-1$   produces the    factor of $N^2$ in the planar limit. 
}
\begin{align}\la{47}
&\qquad \qquad \llangle  \P_a(\tau_1)  \P_b(\tau_2)\rrangle 
=\delta_{ab}  \frac{C_0(\l) }{| 2\sin\frac{\tau_{12}}{2}   |^{2}}\  
\ ,  \  \qquad \qquad     C_0 = 2 B(\l)  \ , \\ 
&\qquad \qquad  B(\lambda)\equiv  {1 \ov 2 \pi^2} {d \ov d \log \l}   \lan \WW \ran = 
\frac{\sqrt{\lambda } \, I_2(\sqrt{\lambda })}{4\pi ^2\,  I_1(\sqrt{\lambda })}\ , \la{471}\\
&C_0 (\l\ll 1)= {\l\ov 8 \pi^2} -  {\l^2\ov 192 \pi^2}  + \OO(\l^3)\ ,  \qquad 
C_0(\l\gg 1) = {\sql\ov 2 \pi^2} -  {3\ov 4\pi^2}  + \OO({1\ov \sql})  \ . \la{472}
 \end{align}
The operator $\P_6$  which couples to the loop in this  $\z=1$ case, on the other  hand, gets renormalized and its scaling dimension is a non-trivial function of $\lambda$. 
At small $\lambda$ one gets\footnote{Definition of a good conformal operator may require subtraction 
of a non-zero constant one-point function on the circle, which may depend on the regularization scheme.
}
\be \la{48} 
\lla  \P_6 (\t_1) \P_6 (\t_2) \rra 
=  \frac{C(\l) }{| 2\sin\frac{\tau_{12}}{2}  |^{2\Delta}} \ ,  \qquad  
C = {\l\ov   8 \pi^2}  + \OO(\l^2) \ , \qquad 
\Delta =1  +  {\l\ov   4 \pi^2}  + \OO(\l^2) \  . 
\ee 
 Here the anomalous dimension  can be obtained by direct computation \ci{Alday:2007he} or by  taking  the  derivative 
 of the beta-function \rf{111} at the $\z=1$ conformal point  \ci{Polchinski:2011im}  as in  \rf{3}. 
 The leading  term in $C$ is the same as in \rf{47},\rf{472} as it comes  just from the free-theory correlator. 
 At strong coupling  the ``transverse''  scalars 
 $\P_a$   should correspond to massless string coordinates $y_a$ in $S^5$ 
 directions (with $\Delta=\Delta_+ =1$, cf. \rf{2.6})  \ci{Giombi:2017cqn}
  while   $\P_6$  should  correspond \ci{malda} 
   to the 2-particle  world-sheet  state $y_a y_a$ (see section \ref{sec4.2}), 
      with dimension  $\Delta=2\Delta_+ + \OO({1\ov \sqrt{\lambda}})=2 + \OO({1\ov \sqrt{\lambda}})$  \ci{Polchinski:2011im}. 
The  subleading  term in   \be  \la{50} \Delta=2 - {5\ov \sql} +  \OO\big({1\ov (\sql)^2}\big) ,
\ee 
       computed in 
   \ci{Giombi:2017cqn}    has a negative sign   consistent   with  a possibility of    a  smooth 
    interpolation  to the weak-coupling expansion in \rf{48} (see also section \ref{sec4.2}). 

 In the case   of  the standard  WL   with no scalar coupling ($\z=0$)   
 the defect  CFT$_1$   has unbroken $SO(6)$  symmetry  and thus  all 6 scalars 
 have the same  correlators:
 \begin{align}\la{49} 
& \llangle  \P_m \rrangle
=0 \ , \ \ \ \ \ \ \ \ \qquad 
  \llangle  \P_m(\tau_1)  \P_n(\tau_2)  \rrangle 
=\delta_{mn}  \frac{C (\l) }{|2\sin\frac{\tau_{12}}{2} |^{2\D}}\ ,\\  &
 \qquad     C  =  {\l\ov   8 \pi^2}  + \OO(\l^2)    \ , \ \qquad \qquad 
\Delta= 1  -  {\l\ov   8 \pi^2}  + \OO(\l^2) \  . \la{499}
 \end{align}
 Here the leading  free-theory  term in $C$  is  the same as in \rf{472},\rf{48} as it comes  just from the free-theory correlator. 
 The anomalous   dimension  in \rf{499}  found  by direct computation in   \ci{Alday:2007he}   is again the same   as 
  the  derivative of the beta-function \rf{111} at the $\z=0$ conformal point  \ci{Polchinski:2011im}  (see \rf{3}). 
At strong  coupling, \ie\space in the string theory  description 
where the $S^5$ coordinates are  to be  subject to the Neumann boundary
conditions restoring the $O(6)$ symmetry,  one expects     to find   \ci{Alday:2007he}  
\be \la{500}
\Delta=  {5 \ov \sql}    +   \OO\big({1\ov (\sql)^2}\big),
 \ee
  which is consistent with the negative sign of the anomalous 
dimension at weak coupling in \rf{499}, suggesting that it decreases to zero at strong coupling.\foot{It is interesting to notice
 that  the data \rf{50},\rf{500}  about  strong-coupling dimensions of $\P_6$   near $\z=0$  and  near $\z=1$  
is consistent with   the relation \ci{Polchinski:2011im}   $2\D_+   +  2\D_-=2$, i.e.  $  [ {5 \ov \sql}    +   O({1\ov (\sql)^2})] + [ 2 - {5\ov \sql} +  O({1\ov (\sql)^2})]
= 2 +  \OO({1\ov (\sql)^2}).
$ 
Here $2\D_\pm$ are dimensions of perturbations near the two ends of the flow between the 
Dirichlet  and Neumann  b.c.   which may be interpreted as   being 
driven by  the ``double-trace''-like operator constructed out 
of  a  massless  2d scalar   with  strong-coupling dimensions  $\D_+ =1$ and $\D_-=0$
(see  section \ref{sec4.2}). 
}


As a test of our perturbative calculation of the expectation value \rf{3500}  of the generalized WL \rf{0}, let 
us now relate its expansion near the conformal points $\z=0, 1$ to the above expressions for the 2-point functions 
of the $\Phi_6$ operator. 
The expectation value of  $W^{(\z)}$  for the circular contour  ($| \dot x|=1$) 
   expanded  near  $\z=0$    may be written as 
\be\la{410}
\langle W^{(\z)}\rangle =  W^{(0)}\Big[ 1 + \zeta\, \big\llangle\oint d\tau \, \Phi_6(x(\tau))  \big\rrangle
 +\frac{\zeta^{2}}{2}\,\big\llangle
\oint d\tau \, \Phi_{6}(x(\tau) ) \,\oint d\tau' \, \Phi_{6}(x(\tau')) 
\big	\rrangle +\OO(\z^3)  \Big] ,
\ee
where  $\lla...\rra $  is defined as in \rf{7}  but now  for $\z=0$, i.e. 
 with only  the  gauge field  coupling $i\int d \tau  \dot x^\m A_\m$  in the exponent  and 
  the  normalization factor   $W^{(0)} \equiv \langle W^{(0)}\rangle$  has weak-coupling expansion  given  in \rf{3511}. 
The order $\z$ (tadpole) term here  vanishes automatically as in \rf{49} due to  the 
$SO(6)$ symmetry,       
    consistently    with   the conformal invariance.
We may compute the   $\zeta^2$ term here 
\be
\la{4.12}
\langle W^{(\z)}\rangle_{\z^2} = \frac{\z^2}{2}   W^{(0)}  \,\int_{0}^{2\pi}d\tau\,\int_{0}^{2\pi}d\tau'\,
\llangle\Phi_6(\tau)\,\Phi_6(\tau')\rrangle\  ,  
\ee
  directly   using the  conformal  2-point function \rf{49}  with  generic  $C(\l)$  and $\Delta(\l)\equiv 1 + \gamma(\l)$.  
\iffa  \be  \la{999} 
  \llangle  \P_6 (\tau_1)  \P_6 (\tau_2)  \rrangle 
=  \frac{C (\l) }{|2\sin\frac{\tau_{12}}{2} |^{2\D}}\ , \qquad   
    C =  \frac{\l}{8\pi^2}    + \OO(\l^2)    \ , \ \qquad 
\Delta\equiv  1  + \g(\l) = 1 - \frac{\l}{8\pi^2}     + \OO(\l^2) \  .  \ee
 \fi 
Doing the integral over $\tau$  as in \rf{3331}   and then expanding in small $\l$ using \rf{499}  we obtain\foot{This 
 integral is similar to the one in \rf{3.26}   
and thus   can be found by an analytic continuation in $\gamma$. 
Alternatively, we may  use a cutoff regularization,  see Appendix \ref{A}.}
\begin{align}\la{413} 
\langle W^{(\z)} \rangle_{\z^2}  &
= \zeta^{2}  W^{(0)}\,C(\l)\,\frac{\pi^{3/2}\,\Gamma(-\frac{1}{2}-\gamma(\l))}{2^{1+2\gamma(\l)}\,\Gamma(-\gamma(\l))} \no  \\
&= \zeta^{2}  W^{(0)}\,C(\l)\, \pi^2 \gamma (\l) \big[1  + \OO(\gamma^2) \big]
\ =\    - \z^2 \frac{\l^2 }{64\pi^2}+\mc O(\lambda^{3}) \ . 
\end{align}
This  precisely matches the term of order $\l^2 \z^2  $ in \rf{3500}. 
Comparing to the  general  relation \rf{4},  the higher order terms 
in  the anomalous dimension $\gamma(\l)$   can be  absorbed into the relation between $\C$  in \rf{2}  and $C$ in \rf{49}.

Next, let us consider  the expansion of the   WL  \rf{0},\rf{3500}  near the supersymmetric conformal point $\z=1$. 
The term 
of order $\zeta-1$ in this expansion is expected to vanish by conformal symmetry (provided a possible tadpole contribution is suitably 
subtracted),\foot{As  we  have seen above, the dimensional regularization  scheme that leads to \rf{2}  and 
 thus implies the  vanishing of the tadpole at the conformal point    effectively preserves the conformal invariance.
} 
and the term of order $(\zeta-1)^2$ is to be related to the integrated two-point function on the supersymmetric WL
\be\la{4133}
\langle W^{(\z)}\rangle_{(\z-1)^2} = \frac{1}{2}(\z-1)^2\, W^{(1)}  \,\int_{0}^{2\pi}d\tau\,\int_{0}^{2\pi}d\tau'\,
\llangle\Phi_6(\tau)\,\Phi_6(\tau')\rrangle_{\zeta=1}\ .   
\ee
Inserting here the conformal 2-point function (\ref{48}) and we get  the same  integral as in \rf{4.12},(\ref{413}).
 Plugging  in   the  values for 
$C=\frac{\l}{8\pi^2}+O(\l^2)$ and $\gamma = \frac{\l}{4\pi^2}+\OO(\l^2)$  from \rf{48}  we get 
\be\la{3144} 
\langle W^{(\z)}\rangle_{(\z-1)^2} =\frac{\l^2}{32\pi^2}  (\zeta-1)^{2}+\OO(\l^3) \ , 
\ee 
which is indeed in precise agreement with the term of order $(1-\zeta)^2$ in the expansion of \rf{3500} near $\zeta=1$ 
\be \la{444}
\langle W^{(\z)}\rangle =  \langle W^{(1)}\rangle 
\Big\{1  +  {\l^2\ov 32 \pi ^2}   \Big[{(\z-1)^2}{} + {(\z-1)^3}{}+\tfrac{1}{4} (\z-1)^4 \Big]
  + \OO\big(\l^3\big) \Big\}\,.
\ee 
We may also  compare  the  higher order terms in the    small $\z$    or small $(1-\z)$   expansion to  integrated 
higher-point  conformal correlators  of the  $\z=0$ and $\z=1$    CFT's. 
The absence    of the $\z^3$ term (and other $\z^{2n+1}$   terms)   in the expansion   near the 
$\z=0$   is in agreement    with the vanishing of the   odd-point scalar correlators  that  follows from the $\P_m \to -\P_m$ symmetry of the SYM action. 
At the same   time,   the 3-point scalar $\P_6$ correlator  at the $\z=1$ point  is non-trivial  (cf.  also 
\ci{Kim:2017sju,Kim:2017phs}). In general, on the 1/2-BPS circular WL  we should have 
\be \la{445}
\llangle\Phi_6(\tau_1)\,\Phi_6(\tau_2)\Phi_6(\tau_3) \rrangle_{\zeta=1}
=  { C_3 (\l) \ov  |2\sin\frac{\tau_{12}}{2} |^{\D} \  |2\sin\frac{\tau_{23}}{2} |^{\D}\  |2\sin\frac{\tau_{31}}{2} |^{\D}
}\ , 
\ee
where  at weak coupling $\Delta=1 + \gamma(\l) $ is   the same as in  \rf{48}, i.e. $\g= {\l \ov 4 \pi^2} + \OO(\l^2)$, 
 and  we should have   $C_3= c_3 \l^2 + \OO(\l^3)$.
Integrating \rf{445}  using \rf{3.12}  and  then expanding  in small $\l$ we get as in \rf{413},\rf{4133}   
\begin{align}
&\langle W^{(\z)}\rangle_{(\z-1)^3} =  {1\ov 3!} (\z-1)^3\,   \langle W^{(1)}\rangle  
\oint d\tau_1 \,  d\tau_2 \,  d \tau_3\  \llangle\Phi_6(\tau_1)\Phi_6(\tau_2) \Phi_6(\tau_3) \rrangle_{\z=1} \no \\
&\qquad = (\z-1)^3   \langle W^{(1)}\rangle \, C_3 \, 
\frac{\pi^{3/2}\, [\Gamma(-\frac{\g}{2})]^3 \   \Gamma(-\frac{1}{2}- {3\ov 2} \gamma)}{3\cdot  2^{1+3\gamma}\, [\Gamma(-\gamma)]^3} 
= - {8\ov 3}  \pi^2 (\z-1)^3   C_3 \big[1   +   \OO(\l) \big]\ . \la{4455}
\end{align}
Comparing \rf{4455} to \rf{444} we conclude that 
\be  C_3 =  -{3 \l^2 \ov 256 \pi^4} +   \OO(\l^3) \ . \la{435} \ee
The $\z^4 $   term in the expansion  of \rf{3500}   
should     be related  to the integrated value  of the 4-point correlator  of $\P_6$. To  
 $\l^2$   order it is given just by the  product of the two 2-point contributions 
(corresponding to the two  ladder  graphs; the third ordering  is subleading  in the planar limit) 
 \begin{align} \la{4445}
&  \llangle \Phi_6(\tau_1) \Phi_6(\tau_2)\Phi_6(\tau_3)\Phi_6(\tau_4) \rrangle  = \big[ G_0(\tau_1,\tau_2) \,  G_0(\tau_3,\tau_4) + G_0(\tau_1,\tau_4) \,  G_0(\tau_2,\tau_3)
\no \\    &  \qquad \qquad\qquad \qquad\qquad \qquad\qquad \qquad
 + \OO(\l^3) \big] \theta(1,2,3,4) + {\rm permutations}  \ , \end{align}
where  $G_0(\tau_1,\tau_2)= {\l\ov 8 \pi^2}  \frac{1 }{|2\sin\frac{\tau_{12}}{2} |^{2}}$ is the 
leading term in the 2-point correlator \rf{49}  of $\P_6$ at $\z=0$  and 
$\theta(1,2,3,4)= \theta(\tau_1-\tau_2) \theta(\tau_2-\tau_3)\theta(\tau_3-\tau_4)$. 

To understand the precise relation  between the integrated  4-point correlator and the  $\z^4$ term in $\langle \WZ\rangle$ in \rf{3500} 
one should  follow the logic of  conformal perturbation  theory 
by a nearly-marginal  operator $O$  with dimension $\Delta=d-\eps$ (see, \eg, \ci{Fei:2015oha}). In the 
present case  of  $d=1$   near the $\z=0$ point  we have $O=\P_6$  with dimension $\Delta=1 -\eps, \  \eps\equiv -\gamma = {\l\ov 8\pi^2}+... \ll 1$ (see \rf{499}). Then the  dimension 1  perturbation   $\z_\bb O$ where the  bare  coupling $\z_\bb$ 
is related to the dimensionless  renormalized   one  by 
$\z_{\bb} = \mu^{\eps} ( \z + {\l \ov 16\pi^2   \eps }  \z^3 +...)$   corresponding  to the  beta-function \rf{111}, i.e. 
$
\b_\z = -\eps\, \z + {\l\ov 8\pi^2} \z^3+... $. 
Computing $\langle \WZ\rangle$ in an expansion in powers of $\z_b$   we get  for the $\l^2$ term:  $
\langle \WZ\rangle= \langle W^{(0)}\rangle \big[1+ \l^2( k_2 \z_\bb^2 +  k_4 \z_\bb^4) + \OO(\z_\bb^6)\big]$ where
$k_2 = - {1\ov 64\pi^2}$  is  the contribution of the  integrated 2-point function 
given by \rf{413}  and $k_4 = {1\ov 64\pi^2} ( \pi^2 + \ha \pi^2) $ 
is  the  contribution of the integral of \rf{4445}, \ie\space  
the sum of the $\z^4$ terms in  the two ladder diagrams in \rf{3556}.
Similarly  to   what happened in  the dimensional  regularization  case  in \rf{2633}, 
here the quadratic  term contributes to the quartic one once expressed  in terms of the   renormalized coupling.  Using 
$\z_\bb=\z+ \ha \z^3 + ...$  we get 
$k_2 \z_\bb^2 +  k_4 \z_\bb^4 = k_2 \z^2  + k_4' \z^4+ ...$,  where $k_4'=k_4  + k_2= {1\ov 128\pi^2}$ which 
is  in agreement with  the $\z^4$ coefficient in \rf{3500}.
Similar   considerations   should apply to the $(\z-1)^4 \l^2$ term in the expansion \rf{444}  near $\z=1$.

\iffa  \footnote{
\red{Potential issue:} if we assume that the 4-point function 
$G_{4}(\bm{\tau}) = 
\llangle\Phi_6(\tau_1)\,\Phi_6(\tau_2)\Phi_6(\tau_3)\,\Phi_6(\tau_4)\rrangle$ is symmetric under
all permutations 
$\tau_{i}\leftrightarrow \tau_{j}$ as a consequence of crossing symmetry, then the relevant integral contributing 
to $\langle W^{(\z)}\rangle_{\zeta^{4}}$ is the unrestricted integral $\int_{0}^{2\pi}
d^{4}\bm\tau\,G_{4}(\bm \tau)$ and each factorized contribution is proportional to the 
square  $(\int d\tau \llangle\Phi_{6}(\tau)\Phi_{6}(0)\rrangle)^{2}$ that vanishes at tree level,
{\em} cf. for instance (\ref{413}) that is zero for $\gamma\to 0$. This argument is not valid
because crossing symmetry of $G_{4}(\bm{\tau})$ is not manifest at weak coupling where the 
planarity constraint forbids the crossed exchange diagram. The situation is different at strong coupling
where all three channels contribute, see for instance the discussion in footnote 9 of 
\cite{Giombi:2017cqn}. \red{A possibly related fact} is the remark that the sum of all channels,
so including the non planar one, is zero at lowest order in 4 dimensions, see (6.3) and (6.4) 
of \cite{Bianchi:2013rma}.
}
\fi 

 \iffa  
 for operator  like $(\Phi)^J$, e.g., $J=1$ for $\Phi^I$.. 
At weak coupling for $J=1$:    $\Delta=1 - {\l \ov 8 \pi} + O(\l^2)$. 
This is   like BMN: expansion near  geodesic in $S^5$  (com or zero mode quantization) 
 or rather   dim of 
 vertex operator $e^{i  J\vp }$  as in open string in flat space  of $- {1\ov \sql} \nabla^2$   operator value  as suggested in \ci{Polchinski:2011im}.
 This is then different pattern from the one suggested in \ci{Giombi:2017cqn}  for operators on WML where 
 $\Phi^i$ was directly  represented by string  coordinates $y^i$. 
 In WL  case   \adst theory should have $O(6)$ invariance  and susy   should be   broken. 
 $\Delta=0$   for  $S^5$   scalars  suggests that we need    vertex operators or derivatives  of them -- integrate over zero mode... 
may be    in WL case with $\Delta=0$ better use   $\del y^i$ as operators  with dim 1  -- 
again interpolation  to scalars  at weak coupling. 
generalization to z(t) ?   funct deriv 
\fi 

\section{Strong coupling   expansion }\label{secstr}

As   discussed    in \ci{Alday:2007he,Polchinski:2011im}, the \adss   string description of the standard Wilson loop 
  should be  given by the  path integral  with Dirichlet boundary condition along the boundary of $AdS_5$   and Neumann 
    (instead of Dirichlet for the Wilson-Maldacena loop)  
condition for the $S^5$ coordinates. The case of the   generalized WL \rf{0}  
may then correspond  to mixed boundary   conditions \ci{Polchinski:2011im}.
Below we shall first   discuss   the subleading strong-coupling correction  to  the standard   WL  ($\z=0$)   comparing it   to the 
more familiar    WML   ($\z=1$)   case  and   then  consider the  general  $\z$  case. 

The  strong  coupling expansion of 
  the straight-line  or circular  WL   will    be represented by the string partition function  with 
the same   \adst   world-sheet  geometry  as in  the WML  case \ci{Berenstein:1998ij}.  
%
 As  the  \adst   is a  homogeneous-space, 
the  log of  the  string partition function 
should   be proportional to  the   volume of \adst \cite{Drukker:2000ep,Buchbinder:2014nia}.
In the  straight-line case  the 
volume of \adst  with infinite ($L\to \infty$)  line as a boundary is  linearly  divergent as 
 ${L\ov a}$.
Thus the  straight-line  WL    should be     given
just by an   exponent of this  linear  2d  IR   divergence.    
Linear  UV divergences in WL for a smooth contour are
known to  factorize  in general  at   weak coupling  \cite{Dotsenko:1979wb}.
After the  separation of this linear divergence   the  straight-line WL     should be   thus  equal to 
  1  as   in the case  of   the  locally-supersymmetric   WML.
The same should  be true for  the generalized  WL \rf{0}. 

Similar arguments  apply in the case of the circular WL  where  the minimal surface is
again the \adst but  now with a  circle as its boundary.  In this case  the volume  is   (we fix the radius to be 1)
\be \la{2.1}
V_{\rm AdS_2} = 2 \pi \Big( {1\ov a} - 1\Big)  \ , \ee
 i.e.  has  a finite part    and thus  
 the expectation value    may be a non-trivial  
   function of string tension $\sql\ov 2\pi $.  After factorizing  the linearly divergent factor, 
   the leading  strong-coupling   term   will then have a   universal $\sql$  form 
   \be \la{21} 
   \lan W^{(\z)}  \ran \equiv e^{-F^{(\z)} (\l) }  \  , \qquad \qquad   F^{(\z)}  =-\sql + F^{(\z)}_1 + \OO( \tfrac{1}{\sql})    \ . \ee
   The subleading terms $F^{(\z)}_1+ ...$    will, however, 
    differ due to the  different boundary conditions  in the $S^5$ directions.
    
    \iffa \foot{
      puzzle with strong-coupling claim that   < WL>  ~  exp   sqrt lambda    just 
like for BPS one. 
in BPS case tr P exp   i A +   P   is not unitary matrix due to scalar coupling so it may not be  surprising that   at strong coupling   < W >   >>1    instead of < 1. 
But   for standard WL    we   expect  |< W >|  <1   always?   
The resolution is   "no"     due to linear divergences -- 
exp (-  sql  ( 1/eps -1)  )   indeed  < 1    for any finite cutoff; it is the renormalized  value that behaves strangely, but there is no paradox. 
    } 
   \fi 
 
\subsection{Standard Wilson loop}\la{sec4.1}

   Let us   consider    the  1-loop string correction in the standard WL  
    case  following the same   approach as used in the WML case in 
   \cite{Drukker:2000ep,Kruczenski:2008zk,Buchbinder:2014nia}.
 As the  fluctuation  determinants for all the 2d  fields (3     AdS$_5$ bosons  with $m^2$=2, 8  fermions  with $m^2=1$   and ghosts) 
   except the   $S^{5}$    massless
scalars  are the  same,   
  the ratio of the WML  and WL   expectation values \rf{1111}    should be 
 proportional   the  ratio of the   1-loop 
string partition functions with the  Dirichlet  
 and Neumann   boundary conditions  in the five  $S^5$ directions:
\be\la{2.2}
\frac{\langle \WW\rangle} {\langle \W \rangle} =  {  e^{ -  F^{(1)} } \ov e^{- F^{(0)}}} =  \  \N_0^{-1} \  
\Big[\frac{\det (-\na^{2})_{\rm D}}{
\det' (-\na^{2})_{\rm N}}\Big]^{-5/2}\, \Big[1+\mc O({1\ov \sqrt\lambda})\Big] \ . 
\ee
Here  $-\na^{2}$ is the  massless scalar wave operator in AdS$_{2}$   and $\N_0$  
is the  normalization
factor  of the  $S^5$  zero  modes   present in  the  Neumann      case 
\be 
\N_0 =  c_0\,   (\sql )^{5/2}  \ ,  \la{232}
\ee
with   $c_0$    being  a numerical constant (representing   contributions  of renormalized  volume of \adst and volume of $S^5$).\foot{In general,   one is  to separate  the 0-mode
  integral and treat it exactly
  (cf.    \cite{Miramontes:1990nf}).} 
The  1-loop corrections to $F^{(\z)}$ are  thus  related by 
\be
\la{2.3}
 F^{(1)}_{1}- F^{(0)}_{1} = 5\,\Big[
\tfrac{1}{2}\log \det (-\na^{2})_{\rm D}-\tfrac{1}{2}\log{\rm det}' (-\na^{2})_{\rm N}\Big]  + \log \N_0 \ . 
\ee
  To compute this correction we may   use the    general result  
for the difference of  effective actions  with standard (D or +) and alternate (N or -)   boundary conditions for  a  scalar with mass $m$ in  AdS$_{d+1}$  \cite{Hartman:2006dy,Diaz:2007an}
\begin{align}   
\delta \Gamma &= \G_+ -\G_- =
\tfrac{1}{2}\log\det (-\na^2+m^{2})_{\rm D}
-\tfrac{1}{2}\log\det (-\na^2+m^{2})_{\rm N} \  \no
\\
\la{2.5}
 &  = \tfrac{1}{2}\sum_{\ell=0}^{\infty} c_{d,\ell}\,\log
\frac{\Gamma(\ell+\frac{d}{2}-\nu)}{\Gamma(\ell+\frac{d}{2}+\nu)},\qquad \qquad 
c_{d, \ell} = (2\ell+d-1)\,\frac{(\ell+d-2)!}{\ell!(d-1)!} \ , 
\end{align}
where $\nu$ is  defined by 
\be \la{2.6}
m^2 = \Delta(\Delta-d) \ , \ \ \ \ \ \ 
\Delta_{\pm} = \frac{d}{2}\pm\nu, \qquad \nu\equiv \sqrt{\frac{d^{2}}{4}+m^{2}} \ . 
\ee
In the present  case of $d=1$ and $m=0$ 
the $\ell=0$  term   with $\G(\ell+\frac{d}{2}-\nu)$ is singular   and  
should be dropped: this corresponds to projecting out the constant 0-mode present in the Neumann case.
Then in  the limit  $d\to 1$ and $\nu\to {1\ov 2} $ in (\ref{2.5})   we get  (projecting out 0-mode)
\be
\la{2.7}
\delta\G' = -\sum_{\ell=1}^{\infty}\log \ell = 
 \lim_{s\to 0}\frac{d}{ds}\sum_{\ell=1}^{\infty}\ell^{-s} = \zeta_{\rm R}'(0) = -\tfrac{1}{2}
\log(2\pi) \ . 
\ee
One may also   give an  alternative derivation of \rf{2.7}  using the relation between the 
AdS$_{d+1}$ bulk  field  and  $S^d$ boundary  conformal  field partition functions:
$  {Z_{-}}/{Z_{+}} = Z_{\rm conf}$  (see \ci{Diaz:2007an,Giombi:2013yva,Beccaria:2014jxa}).
 For a  massive scalar in  AdS$_{d+1}$   
 associated
 to  an operator with   dimension  $\Delta_+$, the boundary  conformal (source) field has
  canonical dimension $\Delta_- = d - \Delta_{+}$
 and thus the kinetic term $\int d^d x \, \varphi (-\del^2)^\nu  \varphi$, with $\nu=  \Delta_+  - {d\ov 2} $.
In  the present case of the   massless scalar in \adst  we   have 
 $d=1$,   $\Delta_+ = 1, \ \Delta_-=0$  and   $\nu= {1\ov 2} $.  The   induced boundary CFT 
     has thus the  kinetic operator $\del\equiv (-\del^2)^{1/2} $ defined   on $S^1$  and  thus we find again \rf{2.7} 
 \be 
\la{2.9}
\delta\G' = -\log \frac{Z_{+}}{Z_{-}}
= - { 1\ov 4}  \log  {\rm det}'  ( - \partial ^2)  
= -\sum_{\ell=1}^{\infty}\log \ell \  ,
\ee 
where we  fixed the normalization constant in the  $S^1$ eigen-value  to be  1. 

It is interesting to  note  that the zero-mode  contribution   in \rf{2.3} 
may be included  automatically  by ``regularizing''  the $m\to 0$ or $\nu\to \ha$  limit  in \rf{2.5},\rf{2.6}.
One may expect that  for  the Neumann   boundary conditions   which are non-supersymmetric  in the world-sheet theory 
 \ci{Polchinski:2011im}   
the massless  $S^5$  scalars $y^a$    may get  1-loop  correction to  their mass 
$m^2 = - { k\ov \sql}  + \OO( { 1 \ov (\sql)^2}) \to 0$.\foot{This  correction may be 
  found by computing the 1-loop   contribution to  the propagator  of $y^a$ 
 in \adst    background.   Similar   correction  to  scalar propagator  with alternate b.c.   should appear in higher spin theories in  the context of  vectorial AdS/CFT  (there the  effective  coupling is $1/N$ instead of $1/\sql$), see e.g.   \ci{Giombi:2017hpr}. 
 Note that  having a correction to the mass of a  world-sheet   excitation  here does not run against  the  usual 2d 
 conformal invariance  constraint   as we are expanding near a non-trivial background   and are effectively in a 
 physical gauge where the conformal  freedom  is fixed  (cf. \ci{Giombi:2010bj}). 
}
 Then  $\nu = \ha  -  { k\ov \sql}   + ...$  and $\Delta_- =  { k\ov \sql}  + ...$;
  for the  agreement   with \rf{500}   we  need  to fix   $k=5$. 
 We then get an extra $ -\ha \log |m^2| = - \ha \log  { k\ov \sql} $ term  from   the $\ell=0$ term in \rf{2.5}, i.e. 
%
%
%
 \be 
 \delta \Gamma =  \delta \Gamma'  - \ha  \log |m^2| =   -\tfrac{1}{2}
\log(2\pi)  +  \ha  \log \sql - \ha \log k 
     \  .   \la{2111}\ee
     This agrees  with \rf{2.3},\rf{2.7}  if we  set $c_0 = k^{-5/2}$.
 Finally,    from \rf{2.3},\rf{2.5}   we   find 
\be
\la{2.11} F^{(0)}_{1}= F^{(1)}_{1} -  5\delta\G =  F^{(1)}_{1} - 
   5\,\delta\G'   - \log \N_0 =  F^{(1)}_{1}  + \tfrac{5}{2}\, \log (2 \pi)   
      - \big(\tfrac{5}{2}\, \log \sql   + \log c_0\big)    \ . 
\ee

Let us now   recall  that the direct    computation   of the determinants   in 
the  string 1-loop partition function for the circular    WML  
  gives  (after using \rf{2.1}   and separating out the  linear divergence)
\cite{Drukker:2000ep,Kruczenski:2008zk,Buchbinder:2014nia}
(see  also \ci{Kristjansen:2012nz,Bergamin:2015vxa,Forini:2017whz})
\be
\la{2.10}
F_{1}^{(1)} = \tfrac{1}{2}\log(2\pi)\ .
\ee
At the same   time, the exact gauge-theory result  \rf{1.1}  for the WML   implies that 
the total correction to the leading strong-coupling term  should, in fact,  be 
\be
\la{2.12}
 F_{1\, \rm tot}^{(1)} =\  \tfrac{1}{2}\log(2\pi)   - \log 2 +\tfrac{3}{2}\log\sqrt\lambda\ .
\ee
The ${3\ov 2} \log\sqrt\lambda$   term  may be  attributed to   the  normalization of the three  
M\"obius symmetry  zero modes on the disc 
\ci{Drukker:2000rr}, but   the remaining  $\log 2$  difference still remains to be understood.

It is   then  natural to  conjecture  that  for   the standard WL  expanded at strong coupling 
  the  total  value   of the   subleading term  at strong coupling  should   be given by \rf{2.11} 
where the first term   is replaced  by  \rf{2.12}, \ie      
\be
\la{2.13}
F_{1 \, \rm tot}^{(0)} 
= F_{1\, \rm tot}^{(1)}     +    \tfrac{5}{2}\log(2\pi)   + \log \N_0
=   \  3 \log(2\pi)     -   \log  (2 c_0) - \log\sqrt\lambda     \ .
\ee
We  then conclude that   while the leading     $\l \gg 1$  prediction for 
the   log of  the  expectation value  $\tilde F^{(\z)} \equiv  \log \langle \WZ \rangle 
 =  - F^{(\z)}_{\rm tot} $ 
for the 
circular WML  and WL  
   is the same $ {\sql}$ in \rf{21},  
the subleading term in $\tilde F^{(0)}$  is   larger  than that in $\tilde F^{(1)}$  by  $ \log \N_0 = {5\ov 2} \log\sqrt\lambda +...$. 
This   appears to be   in agreement with  a similar   behavior  \rf{190}  observed   at weak coupling  and thus with
 the 1d analog of the F-theorem    \rf{110}.   

While  the  strong-coupling behaviour of WML $ \langle \WW \rangle \sim  (\sql)^{-3/2} e^{\sql} + ... $
follows  from the exact Bessel function   expression  in \rf{1.1},   one may wonder   which  special function may give 
the  above strong-coupling asymptotics  $ \langle \W \rangle \sim  \sql \,  e^{\sql} + ... $   of the standard  WL.


\subsection{General  case}\la{sec4.2}

Turning to the case of  generic $0 < \z < 1$, 
  one may imagine  computing $\lan \WZ(\l)\ran $  exactly    to all  orders  in the weak-coupling 
 expansion  and expressing it in terms of the renormalized  coupling $\z$   (in some particular scheme).
 One  may  then  re-expand   the resulting   function at strong coupling (as in  \rf{1.1})   
 expecting to match      $F^{(\z)}_1$  in \rf{21}   
   with \rf{2.12}   and \rf{2.13}  at the two conformal points. 
 
A  way to set up the strong-coupling (string-theory)  computation  for an arbitrary
  value of  $\z$   may   not  be a priori    clear  as 
  non-conformal WL operators   need not have a  simple  string-theory   description.  
Below we shall  develop  a heuristic  but   rather compelling suggestion  of \ci{Polchinski:2011im}.
Starting with   the \adss   string action  and considering a minimal surface ending, e.g., on a line   at the boundary of AdS$_5$ 
we may choose a static  string gauge where
 $x^0=\tau, \ z= \s$ so that the induced metric is  the \adst one: \  $ds^2= 
{1\ov \s^2} ( d\s^2 +d\tau^2)$; in what follows we identify $z$ and $\s$.\foot{
In the case of the circular  boundary  $ds^2= {1\ov \sinh^2 \s} (d\s^2 +  d\tau^2)$.}   
Let  the 5   independent $S^5$ coordinates  be  $y^a$  (with  the embedding coordinates  being, e.g.,  
$Y_a= { y_a\ov 1 + {1\ov 4} y^2}, \ Y_6= { 1 - {1\ov 4} y^2\ov 1 + {1\ov 4} y^2}$, $ds^2_{S^5} = {dy_a dy_a \ov (1 + {1\ov 4} y^2)^2}$).
In the WL 
case they are subject to the Neumann   condition $\del_z y^a\big|_{z\to 0} =0 $.  
One may then  start with this Neumann  (\ie\  standard  WL)   case   and 
perturb the   corresponding string action $I^{(0)}$  by a boundary term that should induced the flow towards the 
other (Dirichlet or WML) fixed point
\begin{align}
 \la{4a} 
& I{(\ff)} =   I^{(0)}   + \delta I \ ,  \ \ \ \ \ 
  I^{(0)} = T \int d\tau d z \big( \ha \sqrt h h^{pq} \del_p y^a \del_q y^a + ...\big) \ , \  \ \ \qquad T= {\sql \ov 2\pi} 
 \ ,  \\
&  \delta I=  -  \ff\, T \int d\tau  \,  Y_6   \ ,   \ \ \ \ \qquad     Y_6 = \sqrt{1- Y_a Y_a}= 1 - \ha y_a y_a  + ...   
\la{4aa} \ . 
\end{align}
 In $  I^{(0)}$  we give only the part  depending  quadratically 
 on  $S^5$ coordinates  and 
  $h_{mn}$   is the induced \adst metric.

 Here $\kk$ is a new coupling constant  
 which  should be a strong-coupling counterpart of $\z$:  $\ff=0$ should correspond to $\z=0$   and $\ff=\infty$ to $\z=1$. 
 $Y_6$  is then the counterpart of  the  operator $\P_6$  in \rf{0}   perturbing the $\z=0$ conformal point at weak coupling. 
 
 Note that  for  the  \adst metric  $ds^2= z^{-2} ( dz^2  + d \tau^2)$  with  the boundary at $z=a \to 0$ 
 the boundary  metric  is $ds= a^{-1}   d \tau $  and thus  it may be more natural   to write $\delta I$ in \rf{4aa} as 
 $\delta I=  -  \k\, T \int ds  \,  Y_6$   so that      $ \kk= a^{-1} \k$. 
 Then $\k$ will  always appear together   with   the 
 \adst  IR   cutoff factor $a^{-1}$  which, on the other hand,   can   be  
 also  interpreted  -- from the  world-sheet 
 theory point of view  --   as  playing the same   role as a  UV cutoff $\L$.
 \iffa 
 \foot{
 In   2d theory in \adst   the  boundary UV   divergences   
 will be correlated with the IR divergences: the covariant UV cutoff  will be coupled to 
  conformal factor $\sqrt h = e^{2\rho} $  of  2d  metric  via  $ e^{2\rho} \delta \sigma_p \delta \sigma_p \geq \L^{-2}$  as $\L e^\rho= \L  z^{-1} $.} 
 \fi 
  
The variation  of the action $  I{(\ff)}$    implies  that to linear order in    $y^a$   it should 
 satisfy  the  massless   wave  equation in \adst 
(so that near the \adst  world-sheet 
boundary   $y^a= z^{\Delta_+}  u^a  + z^{\Delta_-}  v^a+ \OO(z^2) = z\, u^a + v^a  + \OO(z^2) $)
     subject to   the  mixed (Robin)  boundary condition\foot{The tangent vector  to the boundary is $t^p=( 0,  z)$ and 
 the   outward normal to the boundary is  $n^p= (-z, 0)$,  so that  $h^{pq} = n^p n^q + t^p t^q$.}  
\be \la{552}
 (-\del_z + \ff) y^a  \big|_{z\to 0} =0\ , \ \ \ \   {\rm i.e.} \qquad  - u^a  + \ff\,  v^a =0 \ . 
  \ee 
The parameter $ 0 \leq  \ff \leq \infty$ thus   interpolates between the Neumann and Dirichlet   boundary 
conditions. 
Note that  in general one may add,  instead of $Y_6$,   in \rf{4aa}   any  linear combination $\theta^m Y_m $  with $\theta_m^2=1$ (cf. \rf{0})
and the $S^5$  part of  \rf{4a} as   $\del^p Y^m \del_p Y_m$,  with $Y^m Y_m =1$.  Then the  boundary condition becomes 
$ \big[ -\del_z  Y_m  + \ff (\theta_m - \theta^k Y_k Y_m) \big]  \big|_{z\to 0} =0$. For $\theta_m$ along 6-th axis this  reduces to \rf{552}  
to linear order in $y_a$.

Like $\z$   at weak coupling \rf{111},   the  new boundary coupling $\ff$ will  need to be renormalized, i.e. it  will be 
running  with 2d  UV   scale.\foot{As already mentioned above,    in the present case of  the boundary 
of the \adst    world sheet    being at $z\to 0$ it  is   natural to add  to the  boundary term 
a factor  of $z^{-1} =a^{-1} \to \infty$
that may then    be interpreted  as  playing the same role  as  the world-sheet UV 
 cutoff $\L$; then this running   may be interpreted as a flow with \adst 
 cutoff.}
 In \rf{4aa}     $\kk$   is a renormalized  coupling of  effective 
 mass dimension 1. 
 In  general,   in  the bare action  one should have  
$\delta I_{\bb} = - \L \kk_\bb   T \int d \tau Y_6$,  where 
$\L  \kk_\bb= \mu \kk \big[ 1  +  K({1 \ov \sql}) \log { \L\ov \mu}\big]  + ... $,  with $\L\to \infty$ being a UV cutoff   and 
$\kk_b$ and $\kk$   being   dimensionless.
We   may   choose    the renormalization scale $\m$    to be  fixed as $\m= R^{-1}$  in 
 terms  of the radius $R$    and  set  $R=1$, i.e.   measuring  scales in units of $R$; then 
we may   effectively  treat   $\kk$ as dimensionless.\foot{In the case 
of the circular boundary  the dependence   on   the  radius $R$ that drops out at the conformal points 
remains for  generic  value of $\ff$ or $\z$. One may fix, for example,  $\mu R=1$  as a renormalization condition, 
or rescale $\kk$ by $R$   to make it dimensionless.}

 Dimensionless  
renormalized  $\kk$   should be a non-trivial (scheme-dependent) 
 function  of  the renormalized    dimensionless  parameter $\z$  and  the string tension  or  't Hooft coupling $\l$
 \be \la{4e}   \kk={\fF} (\z; \l) \ , \ \qquad\ \ \   \ \   {\fF} (0; \l) =0 \ , \ \ \    {\fF} (1; \l\gg 1) =\infty \ .
 \ee
  Lack of  information about this function  prevents one from   direct comparison of weak-coupling  and
  strong-coupling pictures.  Just as  an illustration, one may   assume that at large $\l$  one has 
$ {\kk} =  { \z \ov 1-\z} $, ensuring the right limits (cf. \rf{552}).

The   boundary $\kk$-term in \rf{4a}  may be  viewed as a special  case of an ``open-string tachyon''  coupling  depending on $S^5$ coordinates:
\begin{align}\la{44d}
&\delta I_\bb = \L  \int d \tau\, \TT_\bb (y) \ , \ \ \ \ \    \L  \TT_\bb  = \mu \big[ \TT    - \log {\L\ov \mu} ( \a'  D^2 + ...)   \TT + ...\big]\ , 
\\
&\beta_\TT =\mu { d \TT \ov d \mu} =  - \TT  - \a' D^2 \TT + ... \ , \ \ \ \ \ \ \   \te \a' = {R^2\ov \sql} \ . 
\la{44e}
\end{align}
Here $D^2$ is the Laplacian on $S^5$ (of radius $R$ that we set to 1) 
  and $\beta_\TT$   is the corresponding renormalization group function  \ci{Callan:1986ja,Tseytlin:1986tt}.\foot{Similar expression for the  closed-string tachyon  beta-function   has
familiar   extra factors of 2  and $\ha$: 
$\beta_{\rm T} =  -2 {\rm T}   - \ha \a' \nabla^2 {\rm T} + ...$. 
}
 The $\TT= \kk Y_6 $ term in  $I(\kk)$ in \rf{4a} 
 is  the eigen-function of 
 the Laplacian  with eigenvalue $5$  
({\em e.g.}  for  small $y_a$ one has  $D^2  Y_6 =(\del_y^2+ ...)  ( - \ha  y_a y_a  +...)= -5 + ...$).
\foot{
In general,  the eigenfunctions of Laplacian on $S^5$  are 
$C_{m_1 ... m_J}   Y^{m_1} ... Y^{m_J}$  (where  $C_{m_1 ... m_J} $ is totally symmetric and traceless)
with eigenvalue $J(J+ 4)$. For  example,  one may  consider $(Y_1 + i Y_2)^J$. 
In  $J=1$   case  we may choose  any  linear combination $C_m Y^m$ or   any of  six $Y_m$ 
which will  have the eigenvalue $5$.
}
As a result, we should expect to find that   $\kk$   should be renormalized 
 according to 
 \be \la{40c}
\L  \kk_\bb = \mu \kk \big( 1 + {5\ov \sql} \log {\L\ov \m} + ...  \big) \ , \ \ \ \ \ \ 
\beta_\kk = \mu {d \kk \ov d \mu} =\big(  - 1 +  {5\ov \sql} + ...\big)\kk + ... \ . \ee
This  beta-function  then gives  another derivation 
of the strong-coupling dimension \rf{500}
of the perturbing operator   near the  WL  ($\z=0$)  or $\kk=0$   fixed point: 
the coefficient of the linear term in the beta-function should be  the anomalous
  dimension or 
  $\Delta-1$.\foot{  
To recall, the  argument for the strong-coupling dimension  $\Delta(0)  = {5\ov \sql}  + ...$  
of the scalar operator on the WL 
 in \ci{Alday:2007he}
was based on considering \adst in global  coordinates  as conformal to a strip 
$ds^2 = {1 \ov \sin^2 \sigma} ( dt^2 + d \sigma^2) $  where $0 \leq \sigma < \pi$.
Then the Hamiltonian with respect to global time is the dilatation operator and the
mode  constant in $\sigma$   should be the primary, and its energy is the conformal dimension. 
The Hamiltonian of quantized  massless particle moving on $S^5$   is then proportional 
to the Laplacian on $S^5$ 
 with the  eigenvalue ${\alpha' \ov R^2}  J (J+4)$  
 with the present case being that of $J=1$
 (in the $\z=0$   case the  dimension of all 6 scalars  is the same 
 due to unbroken $O(6)$ symmetry).} 
This operator identified as $\P_6$   from the weak-coupling point of view is thus  naturally associated with the 
 quadratic $y_a y_a$  perturbation in \rf{4aa} \ci{malda,Giombi:2017cqn}. 
 
 Note that in  the opposite  WML ($\z=1$) or $\kk\to \infty$ 
 limit we may expect to find the same linear beta-function   but with the opposite coefficient,
   as seen  by rewriting the RG equation in \rf{40c} as 
 $\mu {d \kk^{-1}  \ov d \mu} = - \big(  - 1 +  {5\ov \sql} + ...\big)\kk^{-1}  + ...$, with  now $\kk^{-1}  \to 0$
 (an alternative  is to reverse the UV and IR limits, i.e. $\log \mu \to -\log \mu$).   
 Then the  strong-coupling dimension of $\P_6$  should be   given by $\Delta-1=  1 -  {5\ov \sql} + ...$ in agreement with 
 \rf{50}.

 Another   way to derive \rf{40c}   is  to use 
  the general expression  for the   divergence   of   the determinant  of a  2d scalar  Laplacian    in curved 
 background subject to  the  Robin   boundary condition  $(\del_n + \k) \phi\big|_{\del} =0 $  as in \rf{552} 
  \ci{McKean:1967xf,Kennedy:1979ar} (see also   Appendix B  in \ci{Fradkin:1982ge}   for a review)
\begin{align}  \la{266}
& \G_\infty= \ha \log \det ( -\nabla^2  +  X)\Big|_{\infty}  = - \ha  A_0  \L^{2}   -  A_1  \L   -  A_2 \log\L  
  \ ,  \\
& A_0 = {1 \ov 4 \pi} \int d^2 x \sqrt g  \ , \ \  A_1 = {1\ov 8 \sqrt \pi} \int_{\del} ds   \ , \ \ \ \ \ 
A_2 =     { 1\ov 6} \chi   - {1 \ov 4 \pi} \int d^2 x \sqrt g  X -   {1\ov 2 \pi}    \int_{\del}  ds  \, \k\ .  \no 
\end{align}
Here $\chi$ is the Euler number  and  $L=\int_{\del} ds $ is the length of the boundary. 
In the present  massless case $X=0$ and for  the  Euclidean \adst  we have $\chi=1$.
For the  circular boundary   at $z=a\to 0 $   we have (for $R=1$)   $L=2\pi  a^{-1}$.
To compare   this to \rf{552}   we note that for  an outward normal to the boundary of \adst we have 
$(\del_n + \k) \phi\big|_{\del} = ( - z \del_z + \k) \phi \big|_{z=a} $
so  that  we need  to  identify $ a^{-1} \k$ with $\kk$ in \rf{552}. 
Taking into  account  the factor of 5 for    massless scalars  $y_a$ 
we thus  find the same $\kk  \log \L$   divergence as in \rf{40c}. 
\foot{Note that \rf{266}  directly 
applies only for a  finite non-zero $\kk$ (including $\kk=0$ of  the Neumann condition). In the Dirichlet case  ($\k \to \infty$) 
the sign of $A_1$ is reversed   and  the boundary  contribution  to  the logarithmic divergence (the last term in $A_2$) 
is absent. 
Thus   the D-limit  or $\kk\to \infty$  can not be  taken directly in \rf{266} (see also \ci{Dowker:2004mq}). 
The  logarithmic $\chi$   divergence  and the quadratic   divergence are universal,  
 so they  cancel  in the difference     of effective actions 
with  different   boundary conditions.  Linear  divergence   has the opposite sign for  the Dirichlet and    Neumann   or Robin  b.c.;
 that means it cancels   in the difference of effective actions for the Robin and  the Neumann conditions \rf{2.56}.}

   
 Explicitly, in  the case of  5  massless  scalars  in AdS$_{d+1}$  with  spherical  boundary   and  
 mixed boundary conditions  \rf{552} 
  the analog of \rf{2.5}     gives     \cite{Hartman:2006dy,Diaz:2007an}
 (see eqs.(3.2),(5.2) in  \cite{Diaz:2007an}) 
  \be
\la{2.55}
  F_1{(\ff)}-  F_1{(0)} = 
   \tfrac{5}{2}\sum_{\ell=0}^{\infty} c_{d,\ell}\,\log\big( 1 +    \ff \, q_\ell\big) \ , \qquad \qquad
q_\ell  ={   2 ^{2\nu} \   \Gamma(1+\n) \ov \Gamma( 1 - \n) } 
 \frac{\Gamma(\ell+\frac{d}{2}-\nu)}{\Gamma(\ell+\frac{d}{2}+\nu)} R^{2 \nu} \ , 
\ee
  where   $c_{d, \ell}$   is the same   as in \rf{2.5} and in the present   $d=1$  case      $c_{d,0} = 1, \ c_{d,\ell>0}=2$.
   Since  $\kk$ and $\z$  are related by \rf{4e} 
  the connection to previous notation in \rf{2.3},\rf{2.11},\rf{2.10}  is
  \be \la{5522}   F(\kk) \equiv  F^{(\z)} \ , \qquad 
  F(\infty) \equiv  F^{(1)}  \ , \ \ \ \   \ \ \ \ \   F(0) \equiv  F^{(0)} \ . \ee
  Then  from \rf{2.55} 
     \be \la{2.56}
  F_1{(\ff)}-  F^{(0)}_1 =  5  \sum_{\ell=1}^{\infty}       \log\big( 1 + {\ff}\, {\ell}^{-1} \big)  +  \te  {5\ov 2} \log \big(1  + \kk\, |m^{-2}|\big) \ .
  \ee
  Here we    effectively   set  the radius $R$   to $1$  
  absorbing it into  $\kk$ 
   (which will then be dimensionless) 
  and  isolated    the contribution of the $\ell=0$  mode (using that for $m^2\to 0$  we have  $\nu= \ha +  m^2 + ...$). 
   The limit  $\kk \to 0$   of \rf{2.56} is   smooth  provided it is taken before $m^2\to 0$ one. 
  The limit  $\kk\to \infty$  in \rf{2.56} 
  may be formally taken before the 
 summation  and then (using $  \sum_{\ell=1}^\infty 1 + \ha   =  \zeta_{\rm R}(0) + \ha   = 0$)
  we  recover the previous $\z=1$ result in \rf{2.7},\rf{2111},\rf{2.11}.
 
 Using \rf{2.7},\rf{2111},\rf{2.11}   we  may instead  consider  
 the difference  between $F_1{(\ff)}$   and  $ F_1{(\infty)} \equiv  F^{(1)}_1$, \ie
     \be\la{2156}
  F_1{(\ff)} -  F_1{(\infty)}
  =  5  \sum_{\ell=1}^{\infty}       \log\big(\ell + {\ff}\, \big)   
  +  \te  {5\ov 2} \log \kk  \ ,     \ \ \ \ \  \ \ \ \kk >0 \ . 
  \ee
  where   we assumed $\kk>0$ to drop 1 in the log in the second term in \rf{2.56}   and  observed that 
  the  constant $S^5$ zero mode  contribution $\sim \log |m^2| $ which is present only in the N-case ($\kk=0$)
  then cancels out. 
   An alternative   is   to   rewrite \rf{2156}  in the form  that has regular expansion  near $\kk=\infty$    
   \be 
 F_1{(\kk)}-  F_1(\infty) =    5  \sum_{\ell=1}^{\infty}      \log ( 1 + {\kk}^{-1} \ell)   \ , \la{256x} \ee
      where we used  again  the zeta-function regularization  ($\zeta_{\rm R}(0) + \ha   = 0$).
Note that this expression comes out of the general expression in \ci{Diaz:2007an} or \rf{2.55} 
 if we interchange  the roles   of  $\Delta_+$ and $\Delta_-$ (\ie\   set  $\nu= - \ha$)  and replace $\kk \to \kk^{-1}$. 
The infinite sum in \rf{2156} or \rf{256x}   contains  the expected 
 logarithmic UV   divergence as in \rf{40c},\rf{266}  ($\eps = {\L }^{-1} \to  0$)
 as can be seen using an explicit cutoff, $\sum_{\ell=1}^{\infty}   e^{-   \eps \ell}     \log\big( \ell  + {\kk} \big)  \to    \kk \sum_{\ell=1}^{\infty}   e^{-  \eps \ell}  {\ell}^{-1} + ...  =
 - \kk \log \eps + ... $ (we ignore power divergence as in \rf{2.7}). 
 In general, the term linear in $\kk$ in the finite part  is thus scheme-dependent. 
 The finite part of \rf{256x} can be  found using derivative of the Hurwitz   zeta-function  or  simply  
 expanding  the log   in powers of $\kk^{-1} \ell$ and then using   the zeta-function   to define the sum over $\ell$. As a result, 
  \begin{align}  \la{256y}
      &F_{1\, \rm fin}{(\kk)}-  F_1(\infty) =\te     5 \kk (\log \kk -1)   -  5 \log\big [\Gamma(1+ \kk)\big ]  + {5\ov 2} \log ( 2 \pi \kk)\no    \\
      &\qquad  \qquad = 5  \sum_{n=1}^\infty  {(-1)^n \ov n}\, \zr(-n)\,  \kk^{-n} =    - { 5 \ov 12 \kk}   +  { 1 \ov 72 \kk^3} - {1\ov 252 \kk^5}  +   \OO\big({1\ov \kk^7}\big) \ .      
       \end{align}
        Taken with the opposite sign, \ie\  $ \td F_{1\, \rm fin}{(\kk)}- \td  F_1(\infty) $, 
      this  expression     is a positive monotonically decreasing function   which   is consistent with the F-theorem \rf{190},\rf{110}.

\iffa

   


Using ``proper-time'' regularization   we  find that the infinite sum  in \rf{2.56} has  indeed the expected 
 logarithmic UV   divergence as in \rf{40c},\rf{266}  ($\eps = {\L }^{-1} \to  0$) 
 \begin{align} 
 \G(\kk)&\equiv \sum_{\ell=1}^{\infty}   
   \log ( \ell  + {\kk} )  
   \Big|_{\rm reg} = 
 - \int_{\eps }^\infty  {dt \ov t}    \sum_{\ell=1}^{\infty}   e^{- t(\ell  + \kk ) } 
 = - \int_{\eps  }^\infty  {dt \ov t} \  { e^{-t \kk} \ov e^t-1 }  =  \G_{\infty} + \G_{\rm fin} \ , 
 \no \\
  \G_{\infty} & = -  { \Lambda}   +   \kk \log { \L }  \ , \qquad \qquad 
 \la{4234}
\G_{\rm fin} 
\te 
= - \gamma_E  \kk - \log \big[\Gamma(1+ \kk) \big]   + {1\ov 2} \log ( 2 \pi) \ , 
\end{align}
where in \rf{4234}   we used that $ \int_{0 }^\infty  dt \ { t^{n-1}  \ov 1-e^t } =- \Gamma(n)\, \zr(n) $. We may 
 drop power divergence as in \rf{2.7}.\foot{Equivalently,  $\sum_{\ell=1}^{\infty}   e^{-   \eps \ell}     \log\big( \ell  + {\kk} \big)  \to    \kk \sum_{\ell=1}^{\infty}   e^{-  \eps \ell}  {\ell}^{-1} + ...  =
 - \kk \log \eps + ... $.}  
$\G_{\rm fin}$ has  a  regular expansion at small $\kk$ corresponding to first  expansing log in \rf{2.56} in powers   of $\kk$ and then    summing over $\ell$ using the zeta-function regularization, 
$\G_{\rm fin} = {1\ov 2} \log ( 2 \pi)  - \sum_{n=2}^\infty { (-1)^n\ov  n} \zr (n) \kk^n $.

 In general,  the  coefficient of  the  linear in $\kk$ term  is scheme-dependent; for example, 
defining   the finite  part of $\G(\kk) $ using Hurwitz zeta-function, \ie\ 
as   $-{d\ov d s}  \sum_{\ell=1}^\infty  (\ell + \kk)^{-s} \big|_{s=0}$,
we get $\G_{\rm fin}$    without the $\gamma_E  \kk$ term in \rf{4234}. 
In this scheme, after absorbing    the log divergent term  into the renormalization of $\kk$   in the original action \rf{4a} 
   we end up  with the following  finite 1-loop correction (for $\k >0$)
\be \la{4x}   F_{1\, \rm fin}{(\ff)}-  F^{(1)}_1
\te 
 =  
 -  5 \log\big [\Gamma(1+ \kk)\big ]  + {5\ov 2} \log ( 2 \pi \kk)  \ ,  \ee
   where $F^{(1)}_1$   is given by \rf{2.10}. 
   We observe that near $\k=0$    the difference in \rf{4x}   is negative   so that  reversing the sign of $F$  we get 
   $\td F_1{(\ff\to 0)}-  \td F^{(1)}_1  >0$ 
      which   is consistent with the F-theorem \rf{190},\rf{110}. 
      
   An alternative   is   to   write \rf{2156}  in the form  that has regular expansion  near $\kk=\infty$    
   \be 
 F_1{(\kk)}-  F_1(\infty) =    5  \sum_{\ell=1}^{\infty}      \log ( {\kk}^{-1} \ell + 1)   \ , \la{256x} \ee
      where we used  again  the zeta-function regularization  ($\zeta_{\rm R}(0) + \ha   = 0$).
      Computing  the finite part of this sum  using  derivative of Hurwitz  zeta function  
       and subtracting  a scheme-dependent  linear $5 \kk$ term 
       we   get\foot{This is equivalent  to a scheme   where  one  introduces  a cutoff  on ${\kk}^{-1} \ell$ rather than $\ell$  and that produces an extra $\kk \log \kk$  term.}
      \begin{align}  \la{256y}
      &F_{1\, \rm fin}{(\kk)}-  F_1(\infty) =\te     5 \kk (\log \kk -1)   -  5 \log\big [\Gamma(1+ \kk)\big ]  + {5\ov 2} \log ( 2 \pi \kk)\no    \\
      &\qquad  \qquad = 5  \sum_{n=1}^\infty  {(-1)^n \ov n}\, \zr(-n)\,  \kk^{-n} =    - { 5 \ov 12 \kk}   +  { 1 \ov 72 \kk^3} - {1\ov 252 \kk^5}  +   \OO\big({1\ov \kk^7}\big) \ .      
       \end{align}
       This scheme is simply   equivalent to writing $\log(1+ \kk^{-1} \ell) = -\sum_{n=1}^\infty  {(-1)^n \ov n}\,  \kk^{-n} \, \ell^{n} $ 
        and  then doing the sum over $\ell$ using the Riemann zeta-function regularization. 
        Taken with the opposite sign, \ie\  $ \td F_{1\, \rm fin}{(\kk)}- \td  F_1(\infty) $, 
      this  expression     is a positive monotonically decreasing function. 
      
      \fi

   \iffa 
   
    also worth mentioning that 4.29 in fact can be seen to come from the DD formula reversing the role of \Delta_+ and \Delta_- and interpreting f=1/k.

   We  observe that  
 the     first  derivative  of \rf{4x}   at $\kk=0$ vanishes  in agreement  with  the  expected vanishing of the 1-point function 
     of perturbing operator.   
     The   second derivative   or  $\kk^2$  term should be  related   to the integrated 2-point function of $\P_6$  at strong coupling. 
     
     I agree, end of sect. 4 is still not very satisfactory. Regarding F-theorem, though, I think the log(k) term in the vicinity of k=0 makes F large and negative, so at least in the neighborhood of k=0, \tilde F is positive and large and decreases as k is increased. This \tilde F->infty behavior as k->0 is a bit sutble, probably the correct interpretation is that in the strict k=0 limit the log(k) behavior in F^{(k)} is to be replaced/regularized by the -log(sql) zero mode behavior (sql->infty), as there is no zero mode for any finite k, but it appears for k=0. 

By the way, may be worth mentioning that in 4.28 the finite term linear in kappa is scheme dependent as it depends on the way we subtract the log divergence. For instance if we evaluate the sum in 4.26 say by Hurwitz zeta, d/dz sum_l (l+k)^z |_{z=0}, the final result is 5/2log(2pi)-5 log(Gamma[1+k]) without that linear term. Another way is to take two derivatives w.r.t to k and integrating, which gives A+B k -5 log(Gamma[1+k]) with A,B integration constants that we can try to play with.

\fi 

\section{Concluding remarks}\la{secf}

In this paper we computed the $\l^2$ term in the expectation value of the generalized circular Wilson loop \rf{1}
depending on the parameter $\z$. 
The computation is considerably more  involved than in the  Wilson-Maldacena loop case \ci{Erickson:2000af}. 
In particular, in dimensional regularization, to obtain the finite $\l^2$ part one needs to take into account the ``evanescent'' 
dependence  of the  1-loop term on the  bare  value of $\z$. It would be useful to extend  
the perturbative   computation of $\langle \WZ \rangle$ to $\l^3$ order to see  if  the ladder diagrams 
may still be giving the most relevant contributions, with a hope to sum them up to all orders (at least in the standard WL case). 

The circular loop expectation  value $\langle \WZ \rangle$ admits a natural  interpretation as  a  special $d=1$ case 
of a partition function on  $d$-sphere  and thus satisfies a $d=1$  analog of  F-theorem:
we demonstrated the inequality \rf{190}  at  first subleading orders at both weak and strong coupling.

 \iffa 
  \cob{Map of  operators  to \adst   fields or string coordinates}: 
 
 \bu near $\z=0$:      \ \  O(6) is   unbroken,  
 
  scalars  $\P_m$     $\to $  embedding coordinates of $S^5$ 
 
    $\P_m \    \leftrightarrow  Y_m, $  \ \ \ \ \    $m=1, ...6$ 
 
 \bu   near $\z=1$:  \ \    O(6) is broken by selection of  $\P_6$ direction 
 
 or particular 
 parametrization   of $S^5$ 
 
  $\P_a \    \leftrightarrow  Y_a= y_a + ..., \ \ \ \ \ \ \    \P_6 = 1 - \ha y_a   y_a + ... $  
 
 $\P_a$ and $\P_6$ get different  dimensions 
\fi

 The 2-loop  term \rf{1}  in $\langle \WZ \rangle$ determined in this paper effectively 
encodes several  previously known results about the defect CFT$_1$ defined on the Wilson line: the 
 1-loop  beta-function for $\z$ \ci{Polchinski:2011im} and the related anomalous dimensions of the scalar 
operator $\P_6$ near the  two conformal points $\z=1$ and $\z=0$ \ci{Alday:2007he}. It would be interesting 
to further study the spectrum and correlation functions of operator insertions on the non-supersymmetric ($\z=0$) Wilson line. 
A particularly interesting insertion is the displacement operator $D_i \sim F_{ti}$,  
which has protected dimension $\Delta=2$ as a consequence of conformal symmetry (see e.g. \cite{Billo:2016cpy}). 
The normalization of its two-point correlation function is an important observable of the  CFT, which  should  be a 
non-trivial function of the 't Hooft coupling. This observable is also expected to appear in the small angle expansion of 
the cusp anomalous dimension, or in the expectation value of the WL at second order of small deformations of the loop around the circular shape. 
In the case of the supersymmetric Wilson-Maldacena loop, the analogous observable, known as ``Bremsstrahlung function'', can be determined exactly 
by localization \cite{Correa:2012at} as well as integrability \cite{Correa:2012hh, Drukker:2012de}. It would be very interesting 
to find  the corresponding quantity in the non-supersymmetric Wilson loop case.

 Motivated by the 2-loop expression \rf{1}  one may make a bold conjecture\foot{We  thank R. Roiban  for a discussion of   the possible exact structure  of  $ \langle \WZ \rangle$  and this suggestion.}
  that  to all  orders in $\l$ 
 the renormalized   expression  for  the circular loop  will  depend on $\z$ only through  the combination $(1-\z^2)\l$, \ie\   will  have the form 
 \be 
  \la{000}
  \langle \WZ \rangle =   \WW (\l)   \Big[  1 +  \Z\big( (1-\z^2)\l \big) \Big]  \ , \qquad \qquad
  \Z(x) =  \sum^\infty_{n=2} c_n  x^n 
 \ , 
 \ee
 where  $\WW (\l)  $   is the exact expression for the  WML   given in  \rf{1.1}. 
 If  (in some particular  renormalization scheme)  all $c_n >0$ then for $0 \leq \z \leq 1$ 
 this  function  will 
 have the minimum at $\z=1$  and the maximim at $\z=0$, in agreement with the expected structure of the $\beta$-function  in \rf{2} and  the F-theorem \rf{190}.  
 The  standard WL expectation value will be given by 
 $\W(\l) =  \WW (\l) \big[  1 +  \Z (\l ) \big]$.
  One may  also  try to  determine  the coefficients $c_n$ 
    by  using  that  at each $\l^n$   order   the  term   $\z^{2n}$   with the  highest  power  of $\z$ 
 should  come    from the ladder graphs.  
 The large $\l$  behavior of the WL   in \rf{03},\rf{2.13}    suggests that 
 one should have $\Z(\l \gg 1)  \sim \l^{5/4}$.

While localization  does not apply  to the non-supersymmetric circular Wilson loop case, 
it  would be very interesting to see if $\langle \WZ \rangle$, 
and,  more generally,  the spectrum of local operator insertions on the loop, may be determined exactly  in the planar limit using  the underlying integrability of the large $N$ theory.
   
Another important direction is to understand better the strong-coupling side, in particular, shed light on the  precise correspondence  
between the  ``strong-coupling'' and ``weak-coupling''   parameters $\kk$ and $\z$   in \rf{4e}. 
A related question is about the detailed comparison of the expansion of the Wilson loop expectation value 
near the conformal points to correlation functions of scalar operator insertions at strong coupling.

\iffa 
interpolation,  theory on w-sheet vs open string vertex operator description in O(6) case   etc 
Compute order $\l^2$   terms in scalar 2-point function -- may be not hard? 
Compare to PS beta function -- fix  $\l^2$   term in it ?
That  in WML   $\Phi^6$ at strong coupling has $\D=2$   or should be represented by $y^i y^i$
was implicitly suggested in \ci{Polchinski:2011im}  alread
Dynamical mass generation should again be  relevant ?  $e^{- c\sql} $ corrections  to anom dim ? 
Interpolating function?
string side:  issue of 0-mode integral?  normalization? 
As fermions are massive,  as in the cusp  case we may assume   that $O(6)$ 
invariant sigma model gets mass generation and then may be  there are in addition non-perturbative  corrections due to  non-perturbative mass 
$m \sim \l^{1/8}  e^{- {1\ov 4} \sql} $  \ci{Alday:2007mf}. Then such corrections -- absent in the case of the WML     as there we have Dirichlet bc -- may appear to strong-coupling expansion.  May be this is also  related to the fact that 
WL and WML  results differ   at weak coupling.
$O(6)$ restored for null cusp  -- non-perturbatively?     and there WL=WML  so should be    closely related. 
Exact correction from conf anomaly?   Can it be obtained from integrability? 
Parallel   lines in WL case? 
Note: not all operators in gauge theory   should have regular  image  in   \adss   string theory; it is likely  $\WZ$  does not have   such for any $\z$.
Or:  one will need  to deform  string   action by a boundary term  like one $\int d \tau  Y_6= \int d \tau \sqrt{ 1  -  Y_i Y_i} $   (with coeff depending on $\z$) suggested in 
\ci{Polchinski:2011im}, but that will break  conformal invariance and seems problematic...

\fi

\section*{Acknowledgments}
We are grateful to    L. Griguolo, C. Imbimbo, V. Pestun, R. Roiban, 
 D. Seminara,  D. Young  and K. Zarembo  for   very  useful discussions. 
The work of S.G. is supported in part by the US NSF under Grant No.~PHY-1620542.
AAT thanks KITP  at UC Santa Barbara   for  hospitality while this work was in  progress where his research 
 was supported in part by the National Science Foundation under Grant No. NSF PHY11-25915. 
He  was also  supported by STFC grant ST/P000762/1  
and   the Russian Science Foundation grant 14-42-00047 at Lebedev Institute.


\appendix

\section{Cut-off regularization}
\la{A}

We can  compute the leading order $\l$   contribution to the   Wilson loop in \rf{3.8}  using 
 the explicit UV cutoff $a\to 0$  by   replacing  $x^2$ in the vector field  propagator by $x^{2}+a^{2}$.  
Then  in the line case  we get  (cf. \rf{3.9})
\begin{align}
\la{A.1}
\W_{1} 
&= -\frac{1}{(4\pi)^{2}}\int_{0}^{L} d\tau_{1}\int_{0}^{L}d\tau_{2}\,\frac{1}
{(\tau_{1}-\tau_{2})^{2}+a^{2}}  = -\frac{2}{(4\pi)^{2}}\int_{0}^{L}d\tau\,\frac{L-\tau}{\tau^{2}+a^{2}}\notag\\
&= -\frac{L}{16\,\pi\,a}+\frac{1+\log{(L/a)}}{8\,\pi^{2}}+\mc O(a).
\end{align}
Here for  $L\to \infty$    only the first term  is  relevant  and  this linear divergence  is to be subtracted out. 
\iffa
Here, taking $L\to\infty$ is really problematic. Actually, this logarithmic term must be dropped as it is subleading at large $L$: this situation is different from usual power plus logarithmic UV divergences,
{\em e.g.} in a massive theory where mass is fixed while UV cutoff is large-- there we may subtract 
power divergences and keep logarithms as it sometimes physical. Here, $L$ is an IR cutoff that is also large regardless UV one so only first or  "volume" term is to be kept in (\ref{A.1}).
\fi
For the circle, we have   as in \rf{3.11}  $\W_{1} = \frac{1}{8}+\delta \W_{1}$  where  
\begin{align}\la{a2}
\delta \W_{1} = -\frac{1}{4\pi}\int_{0}^{\pi}\frac{d\tau }{4\sin^{2}\frac{\tau }{2}+a^{2}} = 
-\frac{1}{4\,a\,\sqrt{a^{2}+4}} 
 = -\frac{1}{8\,a}+\frac{a}{64}+\mc O(a^{2}).
\end{align}
The  linear divergence  here is  the same as  in (\ref{A.1})  after  the identification of $L$ with the circle
length $2\pi$.  Subtracting this  linear divergence   we get the same result as in \rf{3.11},\rf{3331}.

The computation of the  integral in \rf{4.12},\rf{413}   can be done   using  similar  cutoff $a$  
\begin{align}
I(a, \gamma) = &\int_{0}^{2\pi}\frac{d \tau }{(4\sin^{2}\frac{\tau}{2}+a^{2})^{1+\gamma}}
= 2\pi \,a^{-2-2 \gamma}\, \,_{2}F_{1}\Big(\frac{1}{2}, 1+\gamma, 1, -\frac{4}{a^{2}}\Big) 
\notag \\
= &\  a^{-2-2\gamma}\,\Big[\frac{\sqrt\pi\,\Gamma(\frac{1}{2}+\gamma)\,a}{\Gamma(1+\gamma)}
+\mc O(a^{3})\Big]+\Big[\frac{\sqrt\pi\,\Gamma(-\frac{1}{2}-\gamma)}{2^{1+2\gamma}\  \Gamma
(-\gamma)}+\mc O(a)\Big]. \la{a3}
\end{align}
Expanding    in  $\gamma  \to  0$,  the  first  term 
  gives just a  power divergence with  no finite $\mc O(a^{0})$ part. The 
leading 
finite part in the second bracket
is  the same as  given  in \rf{413}  found    by directly  computing the integral  in \rf{4.12}
 using an  analytic continuation. 
 Subtracting  power divergence  we get  
$I(a\to 0,\gamma) = \pi\,\gamma +\mc O(\g^{2})$ in agreement with \rf{413}. 
One can check   that the  expansions  $a\to 0$   and $\g\to 0$ here commute.

\section{Computing 2-loop  circle integrals} 
\la{B}

\subsection{Expansion method}

The circle integrals that appear in  the expectation value of the 
circular WL can be computed by using the commonly used expansion method 
(see  \cite{Bianchi:2013rma}, Appendix  B of \cite{Griguolo:2013sma}  and Appendix  G of \cite{Bianchi:2016vvm}). 
Let us first  illustrate  it  on   the example of the 
the  1-loop   integral   in \rf{3.11} or \rf{a3}.  Expanding   power of sine-function   as a series of  exponents 
and setting $\a\equiv  \om -1 = 1-\ve$ we  get\foot{Here we  omit the overall factor
$\G(2-\om)$   that does not contribute to the final result.}   
\begin{align}
\no
\W_{1}(\alpha)  
&= -\frac{1}{16\,\pi^{\alpha+1}}\int^{2\pi}_0 d{\tau_1}d{\tau_2}\ \frac{\cos\tau_{12}}
{(4\,\sin^{2}\frac{\tau_{12}}{2})^{\alpha}}\notag \\
& =  -\frac{1}{16\,\pi^{\alpha+1}}\,e^{-i\,\pi\,\alpha}\sum_{n=0}^{\infty}\binom{-2\alpha}{n} (-1)^{n}  \int^{2\pi}_0 d{\tau_1}d{\tau_2}\ \cos(\tau_{1}-\tau_{2})\, 
e^{i\,(n+\alpha)(\tau_{1}-\tau_{2})} \no \\
& = -\frac{2^{-2 (\alpha +1)} \pi ^{\frac{3}{2}-\alpha } \alpha  \cos (\pi  \alpha )}{\Gamma (2-\alpha ) \Gamma \big(\alpha+\frac{1}{2}\big)} =  \frac{1}{8}+\frac{1}{8}(1 + \log\pi )\,\varepsilon+\OO(\ve^2) \ , \la{b1} 
\end{align}
which is in agreement with \rf{3.11},\rf{3331}. 

At two loops, we need the integrals in \rf{3543};   stripping  off irrelevant factors  these are 
\begin{align}
\la{b2}
\mc W_{2^{(a)}}^{(\zeta)}(\alpha) &= 
\int_{\tau_{1}>\tau_{2}>\tau_{3}>\tau_{4}}d^{4}\bm{\tau}\ 
\frac{(\zeta^{2}-\cos\tau_{12})\,(\zeta^{2}-\cos\tau_{34})}
{(4\,\sin^{2}\frac{\tau_{12}}{2}\ 4\,\sin^{2}\frac{\tau_{34}}{2})^{\alpha}},  \\
\mc W_{2^{(b)}}^{(\zeta)}(\alpha) &= 
\int_{\tau_{1}>\tau_{2}>\tau_{3}>\tau_{4}}d^{4}\bm{\tau}\ 
\frac{(\zeta^{2}-\cos\tau_{14})\,(\zeta^{2}-\cos\tau_{23})}
{(4\,\sin^{2}\frac{\tau_{14}}{2}\ 4\,\sin^{2}\frac{\tau_{23}}{2})^{\alpha}}. \la{bb2}
\end{align}
Applying the expansion procedure as in (\ref{b1}), 
we finally  obtain for $\mc W_{2^{(a)}}^{(\zeta)}(\alpha)$  
\begin{align}
& \mc W^{(\zeta)}_{2^{(a)}}(\alpha) = 
\frac{1}{\pi  (\alpha -1)}\Big\{2^{1-4 \alpha } \zeta ^2 
\Gamma (\tfrac{1}{2}-\alpha )^2 \Gamma
   (\alpha -1)^2 \Big[  \pi ^2 (\alpha -1)^2 [(\alpha -1) \zeta ^2+4 \alpha
   ] \notag \\
   & +[(\alpha -1) \zeta ^2+2] \sin ^2(\pi  \alpha )+(\alpha
   -1)^2 \sin ^2(\pi  \alpha ) \Big((\alpha -1) \zeta ^2 \psi ^{(1)}(1-\alpha
   ) \notag \\
   & +[\zeta ^2-\alpha  (\zeta ^2+4)] \psi ^{(1)}(\alpha
   -1)\Big) \Big]\Big\}\notag \\
   &+\frac{4 \pi ^2 (\alpha -1) \alpha  \Gamma
   (1-2 \alpha )^2 \big[ \pi ^2 (\alpha -1) \alpha  \csc ^2(\pi  \alpha )-(\alpha
   -1) \alpha  \psi ^{(1)}(\alpha -1)-1\big]}{\Gamma (2-\alpha )^4},
   \la{B5x}
\end{align}
where $\psi^{(1)}(z)$ is the derivative of the digamma function. Its expansion around 
$\alpha=1$  gives the expression in \rf{3556}
\be
\la{B6x}
\mc W^{(\zeta)}_{2^{(a)}}(1-\varepsilon) = \frac{\pi ^2 (1-\zeta ^2)}{\varepsilon } 
+ \pi ^2 (3-\zeta ^2) (1-\zeta ^2)+ \frac{\pi^4}{6}+\mc O(\varepsilon)\,.
\ee
For $\mc W^{(\zeta)}_{2^{(b)}}(\alpha)$ a similar  calculation gives 
\begin{align}
\la{BBBxxx}
& \mc W^{(\zeta)}_{2^{(b)}}(\alpha) = 
\frac{2 \pi ^6 \big[   (\alpha -1) \zeta ^2+\alpha \big]^2 \csc ^2(2 \pi  \alpha )}{3 \Gamma (1-\alpha )^2 \Gamma (2-\alpha )^2 \Gamma (2 \alpha )^2}\notag \\
& +\frac{4 \pi ^4 (\alpha -4)^2 (\alpha -3)^2 (\zeta ^2-1) \big(2 \alpha
    [(\alpha -2) \zeta ^2+\alpha -1]+\zeta ^2+1\big)  \csc ^2(2 \pi 
   \alpha ) }{\Gamma (1-\alpha
   )^2 \Gamma (5-\alpha )^2 \Gamma (2 \alpha )^2}\notag \\
   &\times {}_3F_2(1,\alpha ,\alpha ;3-\alpha ,3-\alpha ;1)-\frac{12 \pi ^4 (\zeta ^2-1)^2 \csc ^2(2 \pi  \alpha ) 
   }{\Gamma (4-\alpha )^2
   \Gamma (-\alpha )^2 \Gamma (2 \alpha )^2}\,_3F_2(2,\alpha +1,\alpha +1;4-\alpha ,4-\alpha ;1)\notag \\
& -\frac{32 \pi ^4 (\zeta ^2-1)^2 \csc ^2(2 \pi  \alpha ) }{(\alpha -4)^2 (\alpha
   -3)^2 (\alpha -2)^2 (\alpha -1)^2 (\alpha +1)^2 \Gamma (-\alpha -1)^4 \Gamma
   (2 \alpha +1)^2}\notag \\
& \times{}_3F_2(3,\alpha +2,\alpha +2;5-\alpha ,5-\alpha ;1)
-\frac{4 \pi ^4 (\alpha -1)^2 (\alpha  \zeta ^2+\alpha -\zeta ^2)^2
   \csc ^2(2 \pi  \alpha ) }{\Gamma (1-\alpha )^2 \Gamma (3-\alpha )^2 \Gamma (2 \alpha )^2}\notag \\
   &\times{} _5F_4(1,1,1,\alpha ,\alpha ;2,2,3-\alpha ,3-\alpha ;1).
\end{align}
Expanding this around $\alpha=1$, we obtain the finite  expression given in (\ref{3556})
\be
\la{B8x}
\mc W^{(\zeta)}_{2^{(b)}}(1-\varepsilon) =
\frac{1}{2} \pi ^2 (1-\zeta ^2)^2+\frac{\pi^4}{6}+\mc O(\varepsilon)\,.
\ee

\subsection{Method based on Fourier representation}

In the expansion method, the intermediate calculations are not
manifestly real and cancellation of imaginary parts is often due to non-trivial relations between 
infinite sums. A simpler approach closer to the analysis in momentum space
is based on the Fourier representation of the real even function 
 $(4 \sin^2{ x\ov 2} )^{-\alpha}$
 \begin{align} 
&\frac{1}{(4 \sin^2 { x\ov 2})^{\alpha}} =\frac{1}{2}\aa_0(\alpha)+\sum_{n=1}^{\infty} \aa_n(\alpha) \cos(n x)  \ , 
\label{Fourier-sum}\\
&\aa_n(\alpha)  = 
\frac{1}{\pi}\int_0^{2\pi} dx \frac{\cos(n x)} {(4 \sin^2{x\ov 2} )^{\alpha} }
=\frac{\sec (\pi  \alpha )\ \Gamma (n+\alpha )}{\Gamma (2\,\alpha )\ \Gamma (n-\alpha +1)}
\label{Fourier-coeff}
\end{align} 
Note that near $\a=1$   we have  $\aa_0=\frac{\sec (\pi  \alpha ) \Gamma (\alpha )}{\Gamma (1-\alpha ) \Gamma (2 \alpha )}= \alpha-1+\mc O(\alpha-1)^2$.  

Let   us define the integrals
\begin{equation}
\begin{aligned}
&{\cal I}^{(a)}_{\alpha,\beta} = \int_{\tau_1>\tau_2>\tau_3>\tau_4}d^{4}\bm{\tau}\, 
\frac{1}{(4 \sin^2\frac{\tau_{12}}{2})^{\alpha}\ (4\sin^2\frac{\tau_{34}}{2})^{\beta}}\ , \\
&{\cal I}^{(b)}_{\alpha,\beta}  = \int_{\tau_1>\tau_2>\tau_3>\tau_4}d^{4}\bm{\tau}
\,\frac{1}{(4 \sin^2\frac{\tau_{14}}{2})^{\alpha}\ (4\sin^2\frac{\tau_{23}}{2})^{\beta}}
\label{master}
\end{aligned}
\end{equation}
Using the identity  
\begin{align}
& (\zeta^2  -\cos\tau_{12}) \,(\zeta^2-\cos\tau_{34}) =  4 \sin^2 \tfrac{\tau_{12}}{2} \sin^2 \tfrac{\tau_{34}}{2}
+2\,\big(\sin^2\tfrac{\tau_{12}}{2}+\sin^2\tfrac{\tau_{34}}{2}\big)(\zeta^2-1)+ (\zeta^2-1)^2\, , 
\end{align}
we   find that the   ladder integrals in \rf{bb2}      may be written as 
\begin{align}\la{b11}
\mc W_{2^{(a)}}^{(\zeta)}(\alpha) &= \tfrac{1}{4}\,{\cal I}^{(a)}_{\alpha-1,\alpha-1}
+\tfrac{1}{2}\,\Big({\cal I}^{(a)}_{\alpha-1,\alpha}+{\cal I}^{(a)}_{\alpha,\alpha-1}\Big)(\zeta^2-1)
+ {\cal I}^{(a)}_{\alpha,\alpha}(\zeta^2-1)^2, \\
\mc W_{2^{(b)}}^{(\zeta)}(\alpha) &=  \tfrac{1}{4}\,{\cal I}^{(b)}_{\alpha-1,\alpha-1}
+\tfrac{1}{2}\,\Big({\cal I}^{(b)}_{\alpha-1,\alpha}+{\cal I}^{(b)}_{\alpha,\alpha-1}\Big)(\zeta^2-1)
+ {\cal I}^{(b)}_{\alpha,\alpha}(\zeta^2-1)^2.
\label{Wa-master}
\end{align}
To compute  the integrals (\ref{master}), we use the representation (\ref{Fourier-sum}) and  the integrals
\begin{align}
&\int_{\tau_1>\tau_2>\tau_3>\tau_4}d^{4}\bm{\tau}\, \cos(n \tau_{12})\cos(m\tau_{34}) = 
\begin{cases}0, & m, n >0, \\ 
\frac{2\pi^2}{n^2}, & m=0,n>0, \\ 
\frac{2\pi^2}{m^2}, & n=0,m>0, \\ 
\frac{2\pi^4}{3}, & n=m=0,
\end{cases} \la{b13}
\\
&\int_{\tau_1>\tau_2>\tau_3>\tau_4}d^{4}\bm{\tau}\, \cos(n \tau_{14})\cos(m\tau_{23}) 
= \begin{cases}
0, & m\neq n >0, \\ 
-\frac{\pi^2}{n^2}\,, & m=n >0, \\ 
\frac{2\pi^2}{m^2}\,, & n=0,m>0, \\ 
-\frac{2\pi^2}{n^2}, &m=0,n>0, \\
\frac{2\pi^4}{3}, & n=m=0.\la{b14}
\end{cases}
\end{align}
As a result, 
\begin{align}
&{\cal I}^{(a)}_{\alpha,\beta} =\frac{\pi^4}{6}\aa_0(\alpha)\aa_0(\beta) 
+ \sum_{n=1}^{\infty}\frac{\pi^2}{n^2}\big[\aa_0(\alpha)\aa_n(\beta)+\aa_0(\beta)\aa_n(\alpha)\big]\ ,  \la{b144} \\ 
&{\cal I}^{(b)}_{\alpha,\beta} = \frac{\pi^4}{6}\aa_0(\alpha)\aa_0(\beta) 
+ \sum_{n=1}^{\infty}\frac{\pi^2}{n^2}\big[\aa_0(\alpha)\aa_n(\beta)-\aa_0(\beta)\aa_n(\alpha)-\aa_n(\alpha)\aa_n(\beta)\big] \ . 
\label{master-done}
\end{align}
Plugging  this into \rf{b11} and using the explicit expression of the Fourier coefficients (\ref{Fourier-coeff}), this yields 
\begin{align}
\la{B18x}
&\mc W^{(\zeta)}_{2^{(a)}}(\alpha) = \frac{2 \pi ^6 \big[(\alpha -1) \zeta ^2+\alpha \big]^2 \csc ^2(2 \pi  \alpha )}{3 \Gamma (1-\alpha )^2 \Gamma (2-\alpha )^2 \Gamma (2 \alpha )^2}  
+\sum_{n=1}^{\infty}
\frac{2 \pi ^3 \ \big[(\alpha -1) \zeta ^2+\alpha \big]  \csc (\pi  \alpha ) \sec ^2(\pi  \alpha ) }{n^2 \Gamma (1-\alpha ) \Gamma (2-\alpha ) \Gamma (2 \alpha )^2 \Gamma (n-\alpha +2)} 
 \notag \\
&\qquad\qquad\qquad\qquad \times 
\big[\alpha ^2-\alpha +n^2+\zeta ^2 (\alpha -n-1) (\alpha +n-1)\big] \Gamma (n+\alpha -1).
\end{align}
Evaluating the sum gives  
\begin{align}
\la{B17x}
&\mc W^{(\zeta)}_{2^{(a)}}(\alpha)
= \frac{2 \pi ^6 \big[(\alpha -1) \zeta ^2+\alpha \big]^2 \csc ^2(2 \pi  \alpha )}{3 \Gamma (1-\alpha )^2 \Gamma (2-\alpha )^2 \Gamma (2 \alpha )^2}
+\frac{4^{1-2 \alpha }\pi \big[(\alpha -1) \zeta ^2+\alpha \big] \Gamma (\frac{1}{2}-\alpha)}{\Gamma (1-\alpha )^2}\notag \\
&\times  \Big[\frac{(\zeta ^2-1) \Gamma (\frac{1}{2}-\alpha )}{(\alpha -1)^3} -\frac{\pi ^{3/2} 2^{2 \alpha +1} \big[(\alpha -1) \zeta ^2+\alpha \big]
 \, _4 F_3(1,1,1,\alpha ;2,2,3-\alpha ;1)}{\sin(2\pi\alpha)\Gamma(3-\alpha)\Gamma (2 \alpha )}\Big].
\end{align}
One can check that (\ref{B17x}) is equal to \rf{B5x} by using the identity
\begin{align}\la{b19}
& _4 F_3(1,1,1,\alpha ;2,2,3-\alpha ;1) = \frac{(\alpha -2) \big[\pi ^2 \big(1-6 \csc ^2(\pi  \alpha )\big)+6 \psi
   ^{(1)}(\alpha -1)\big]}{12 (\alpha -1)}.   
\end{align}
%
%
%
Using 
(\ref{master-done}) in (\ref{Wa-master}), we find  in a similar way that  
\begin{align}
\la{B23x}
\mc W^{(\zeta)}_{2^{(b)}}(\alpha) = &
\frac{2 \pi ^6 \big[(\alpha -1) \zeta ^2+\alpha \big]^2 \csc ^2(2 \pi  \alpha )}{3 \Gamma (1-\alpha )^2 \Gamma (2-\alpha )^2 \Gamma (2 \alpha )^2}\notag \\
&-\sum_{n=1}^{\infty} 
\frac{\pi ^2 \big[\alpha ^2-\alpha +n^2+\zeta ^2 (\alpha -n-1) (\alpha +n-1)\big]^2\ \Gamma (n+\alpha -1)^2}
{n^2 \cos^2(\pi  \alpha ) \Gamma (2 \alpha )^2\ \Gamma (n-\alpha +2)^2}.
\end{align}
Evaluating the infinite sum gives  the expression which   is in  agreement with (\ref{BBBxxx}). 
In particular, for $\alpha\to 1$, one finds  again  (\ref{B8x}). 

Let us note  that 
 it is easy to 
extract the $\alpha\to 1$ expansion of  expressions like (\ref{B23x}) without 
computing 
the unwieldy closed form (\ref{BBBxxx}): one   is  to 
separate   the leading contribution at large $n$ in the sum. For instance,
\begin{align}
 & -\sum_{n=1}^{\infty} 
\frac{\pi ^2 \big[\alpha ^2-\alpha +n^2+\zeta ^2 (\alpha -n-1) (\alpha +n-1)\big]^2 \Gamma (n+\alpha -1)^2}
{n^2 \cos^2(\pi  \alpha ) \Gamma (2 \alpha )^2 \Gamma (n-\alpha +2)^2} \no \\
&= -\sum_{n=1}^{\infty}\Big[
\frac{\pi ^2 (\zeta ^2-1)^2 \sec ^2(\pi  \alpha ) n^{4 \alpha
   -4}}{\Gamma (2 \alpha )^2}+\mc O\big((\alpha-1)\,n^{4\alpha-6}\big)\Big] \notag \\ 
&=  -\frac{\pi ^2 (\zeta ^2-1)^2 \sec ^2(\pi  \alpha ) }{\Gamma (2 \alpha )^2}\ \zr (4-4 \alpha
   ) +\mc O(\alpha-1)\ \ \  {\stackrel{\a\to 1}{=}}   \ \ \frac{\pi^{2}}{2}\,(1-\zeta^{2})^{2} \ , \la{b21}
\end{align}
where $\zr$ is the Riemann  zeta-function. 
Adding the $\alpha\to 1$ limit of the first line of (\ref{B23x}), {\em i.e.} $\pi^{4}\ov 6$, we 
reproduce   the expression in  (\ref{B8x}).

\subsection{Alternative approach:  expansion   and summation directly  in $d=4$}
\la{ab3}

The ladder  integrals  in \rf{3543}  were computed using  dimensional   regularization 
with the analytic continuation parameter $\a= \omega-1 = {d \ov 2} -1 = 1-\ve  \to 1$. 
The expansion method and its improved Fourier representation version 
used to compute these integrals 
 involve infinite summations that produce  meromorphic functions of $\a$
that are then evaluated  near  the physical value $\a=1$.
Instead of using analytic continuation in $\a$   one may use a 
  simple  alternative  approach: first   set $\a=1$, use expansion  procedure, do the $\tau$-integrals 
    and then  regularize  the resulting infinite sums. 

For example, starting with $\mc W^{(\zeta)\,(a)}(\alpha)$  in \rf{b2},  setting $\a=1$   and using  \rf{Fourier-sum}, i.e. 
$\frac{1}{4 \sin^2{x \ov 2} } =-\sum_{n=1}^{\infty} n\, \cos(n x)$, and \rf{b13}   we get 
(the same expression is found of course by  setting $\a=1$ in   \rf{B18x})
\begin{align}
\la{B22x}
\mc W^{(\zeta)\,(a)}(1) = 2\,\pi^{2}\,(1-\zeta^{2})\,\sum_{n=1}^{\infty}\frac{1}{n}+\frac{\pi^{4}}{6}.
\end{align}
Comparing to  \rf{B6x}, we see that the pole $1\ov \varepsilon$  there  corresponds to 
the logarithmically divergent sum $2\,\sum_{n\ge 1}{1\ov n}$ in \rf{B22x}. 
The finite parts of  \rf{B6x}   and \rf{B22x} (which are, in general, scheme-dependent) 
 do  not match.  The reason for this  disagreement can be understood as follows. 
 The finite term of  order $\zeta^4$ in \rf{b2} comes from the integral 
\begin{equation}
\begin{aligned}
\int_{\tau_1>\tau_2>\tau_3>\tau_4}d^{4}\bm{\tau}
\frac{1}{(4 \sin^2{\tau_{12}\ov 2}  4\sin^2{\tau_{34}\ov 2})^{\alpha}}=\aa_0(\alpha) \Big[ \frac{\pi^4}{6}\aa_0(\alpha)+ \sum_{n=1}^{\infty} \frac{2\pi^2}{n^2}\aa_n(\alpha)\Big]
=\pi^2+\mc O(\alpha-1)  \la{b24x}
\end{aligned}
\end{equation} 
where we used  that 
\begin{equation}\la{b25}
\sum_{n=1}^{\infty} \frac{2\pi^2}{n^2}\aa_n(\alpha) = \frac{\pi^2}{\alpha-1}+\OO(1)  \,,\qquad\qquad  \aa_0(\alpha)=\alpha-1+\mc O\big((\alpha-1)^2\big).
\end{equation}
The  direct  way  of setting $\alpha=1$ before summation  misses  extra  finite 
$0\ov 0$ term as $\aa_{0}(1)$ is set to zero  from the start.  

This subtlety  does not  appear  in the case of $\mc W^{(\zeta)\,(b)}(\alpha)$   which  does not have a  pole
near $\a=1$ (see \rf{B8x}).  Indeed,  direct evaluation at $\a=1$ (or setting $\a=1$ in 
  \rf{B23x}
before summation) gives,  in agreement with (\ref{B8x})  or  \rf{3556}, 
\be\la{b26}
\mc W^{(\zeta)}_{2^{(b)}}(1) =\frac{\pi^{4}}{6}-\sum_{n=1}^{\infty}
\pi^{2}(1-\zeta^{2})^{2} = \frac{\pi^{4}}{6}
-\pi^{2}(1-\zeta^{2})^{2}\,{\zeta}_{\rm R}(0) = \frac{\pi^{4}}{6}
+\frac{\pi^{2}}{2}\,(1-\zeta^{2})^{2},
\ee
where we used $\zeta$-function  regularization  for the linearly divergent sum.

This  direct procedure  thus gives a vanishing  $\z^4$ 
contribution from the type (a)  ladder diagram integral, i.e. the full $\z^4$ term in the 
  final result  \rf{3500} comes just from the type (b) integral,  avoiding the 
 use of the evanescent  bare coupling  terms   in (\ref{2633}),\rf{269}   required in dimensional regularization. 
 
A weak point of this  regularization method 
 is that it is difficult to apply it  to the self-energy and internal-vertex  diagrams in Fig. \ref{fig:two-loops}
where the Mellin double integral representation is quite useful   when combined 
with dimensional regularization. 
 Nevertheless, it is remarkable that  in this   prescription
the only  finite contribution to   the 2-loop term in \rf{3500}   should come  just from the  ladder  type (b) diagram, 
i.e.  the logarithmically divergent  (and scheme-dependent finite)  parts from other diagrams 
should cancel   against each other.

\bibliography{BT-Biblio}
\bibliographystyle{JHEP}

\end{document}